# On the Compliance of Self-Sovereign Identity with GDPR Principles: A Critical Review

Abubakar Sadiq Shehu
msabubakar-sadiq.it@buk.edu.ng
Centre for Cyber Security, University of Surrey, United Kingdom

*Abstract*— Identity Management Systems (IdMs) have complemented how users are identified, authenticated, and authorised on e-services. Among the methods used for this purpose are traditional IdMs (isolated, centralised and federated) that mostly rely on identity providers (IdPs) to broker trust between a user and service-providers (SPs). An IdP also identifies and authenticates a user on-behalf of the SP, who then determines the authorisation of the user. In these processes, both SP and IdP collect, process or store private users' data, which can be prone to breach. One approach to address the data breach is to relieve the IdP, and return control and storage of personal data to the owner. Self-sovereign identity (SSI) was introduced as an IdM model to reduce the possibility of data breaches by offering control of personal data to the owner. SSI is a decentralised IdM, where the data owner has sovereign control of personal data stored in their digital wallet. Since SSI is an emerging technology, its components and methods require careful evaluation. This paper provides an evolution to IdMs and reviews the state-of-the-art SSI frameworks. We explored articles in the literature that reviewed blockchain solutions for General Data Protection Regulation (GDPR). We systematically searched recent SSI and blockchain proposals, evaluated the compliance of the retrieved documents with the GDPR privacy principles, and discussed their potentials, constraints, and limitations. This work identifies potential research gaps and opportunities.

*Index Terms*—Self-sovereign identity, digital identity management, blockchain, eSSIF, Self-sovereign identity management, GDPR, blockchain, right of data subject, transparency in information processing and exchange, right to be forgotten, revocation and access delegation, eIDAS.

## I. INTRODUCTION

IDENTITY management systems are frameworks built on protocols for access control (identification, authentication, and authorisation) on e-services[81]. They are used to define permissible methods of accessing secured resources by authorised users (persons, institutions and things) who go through an authentication process to prove that they are indeed who they claim to be. Traditional IdMs such as isolated, centralised and federated are based on electronic identity (eID) schemes (e-passport, smart-card, secure log-in, mobile ID and token) for identification and authentication [101]. In an isolated IdM, all registered users with a distinct identifiers or credentials (for example passwords) at an SP are managed and maintained locally by a delegated host SP [10]. As a result, each user is closely linked to a unique set of privileges over local resources. In centralised IdM, users are linked to a single IdP who acts as a trust broker for SPs in a network. Instead of at a single SP, a user authenticates at an IdP in a network of computer nodes. After completing the authentication process, the user can move across the participating network nodes, making requests for services and accessing resources without consciously initiating another authentication procedure. In federated method, IdM is done as a distributed task, by delegating responsibilities, through trust relationships between federated parties, where functional and identity services are separated. A federation is formed between SPs and IdPs, making an SP to trust identities vouched for by IdPs in the federation [26]. Despite simplifying access to e-services, traditional IdMs create more challenges for users by increasing cognitive burden because no single third-party or IdP is capable of maintaining all users digital identity. To reduce this burden, users tend to repeat passwords, sometimes using short insecure passwords, hence exposing them to password vulnerability and other data breaches. It would be catastrophic if an adversary gets access to data residing on IdPs [101]. Other issues with traditional IdMs stem from the IdP's autonomy over the data it manages, which gives it the capacity to excessively watch over users' activities or even pose as a user since they have access to all necessary information. Additionally, all registered users and services below IdP services are affected if IdP services fail. These and other security issues hinder their usability [35]. Third party involvement in access control processes has continued to raise concerns over the security, privacy, confidentiality, and integrity of users' data [36]. To address these issues, governments, businesses, and researchers have been working to strengthen the protection of personal data. In this regard, the California State Legislature introduced the California Consumer Privacy Act of 2018 (CCPA) [24] (data protection laws are sector specific in the USA), the National Information Security Standardisation Technical Committee (NISSTC) in China proposed the Personal Information Security (PIS) [68], and the European Union (EU) rolled out the GDPR [96] among others. Of these laws, GDPR concentrates on protecting personal data of users which can be any information that directly or indirectly identifies a natural person or entity in any scenario (in Europe, data privacy is a basic citizen's right) [100][124]. GDPR supersedes previous data control legislation, as it gives control of personal data back to its owner and legislates higher requirements of access to entities involved in processing users' electronic data. Although the GDPR laws require entities to obtain consent from data owners (data subjects) before processing or sharing their data, nothing prevents these entities from acting otherwise (not seeking consent, processing data even when consent is denied) or acting beyond the consent issued. This necessitates the need for a sovereign, independent concept which does not rely on third parties. Self-sovereign identity is an IdM method that aims to address the trust and data breaches of traditional IdM by giving control of personal data stored in a personal digital wallet to the owner. A digital wallet functions similar to a physical wallet in that it stores all digital credentials which are verified and signed digitally. Verified credentials are quicker to issue and verify than their physical counterparts. Since its introduction, SSI has been implemented in a couple of areas, such as IoT devices [102][76], healthcare systems[27][5], banking and finance [125], education [122][40], among others. SSI reduces the possibility of data breaches, as personal data is less widely available to third parties [100][84].The fundamental building blocks of SSI is made up of the following:

1) Actors (issuer, holder, and verifier)
2) verified credentials (VC)
3) Digital wallet
4) Digital agents
5) Decentralised identifiers (DID)
6) Blockchain/registries
7) Governance (trust) frameworks

In SSI, an end-to-end peer connection is established between a holder and a verifier. A holder uses the VC, stored on their digital wallet to proof their identity directly to the verifier. The verifier is be able to verify the signature on the VC in a blockchain register. Blockchain addresses privacy issues of traditional IdM with distributed nodes that can be used for storing of digital credentials. Also, with blockchain the single-point-of-failure does not exist neither is there an IdP, as data is distributed within the network, giving holders on every node the ability to retain personal identifiable credentials. Since blockchain solutions are emerging technologies in SSI [83], they need to be carefully examined to ensure that,they comply with the seven principles of GDPR: (1) lawfulness, fairness, and transparency; (2) purpose limitation; (3) data minimisation; (4) accuracy; (5) storage limitation; (6) integrity and confidentiality; and (7) accountability. Given the current dominance of blockchain technology to resolving IdM issues, it is not surprising that a number of challenges have emerged. For example, can data owners update their personal data when the need arises or delete a credential or specific attribute.

1) How has blockchain and SSI based systems implemented GDPR compliance in the issuance, ownership and processing of VCs?
2) In compliance with GDPR how are consent, revocation, data erasure and delegation created in blockchain based systems?
3) What are the current research domain that utilise blockchain based technology for delegation?

Although there are review articles on SSI some of which include; [60][50][4][44][41][104][97][8][16][64] [3][38][57], a good number of these works uses a different evaluation approach. To answer the proposed questions, this article concentrates on the following contribution:

1) Fundamental evolution on traditional IdMs and SSI, it's protocols, policies.
2) An analysis of related review works in the literature and comparing them to our work.
3) Presenting a thorough analysis of the literature that uses SSI solutions to fulfill the necessary GDPR principles, along with their strengths and existing challenges.
4) Using the GDPR privacy principles, we assess various SSI works and blockchain based solutions.
5) Extending future SSI research, based on a thorough assessment of current and future needs.

We refer to the owner of data by the terms user, entity, data subject, and holder interchangeably. We also refer to an entity offering a service by the terms entity, service provider, verifier, and service.

This paper is structured as follows: **Section** I introduces the work and provides a general overview of the paper. **Section** II elaborates on the evolution of IdMs, limitations of traditional IdMs and fundamental information on SSI. In **Section** III, we discuss some review works in the literature that relates to the theme of this work and compared them with our work. In **Section** IV, we posed some research questions and provided a systematic review process to address these questions. We analysed how we sourced for articles and criteria for including or excluding these articles, and how selected literature works aim to address the research questions.

For each contribution, we identify and discuss existing challenges and opportunities. In **Section** VI, we discuss GDPR privacy principles and evaluate reviewed works compliance with the privacy principle in **Table** VII. In **Section** VII, we provide further discussion and critical evaluation for the literature works, blockchain and SSI with some limitations and future direction. We conclude this paper in **Section** VIII.

## II. EVOLUTION OF DIGITAL IDENTITY MANAGEMENT

Digital identity is a collection of attributes or identifiers that can be used to uniquely identify a user (fully or partially) in an electronic environment. Digital IdMs is a set of procedures used to store and manage user's identifiers based on access control policies for user identification, authentication, and authorisation. These policies are also used to aggregate users and simplify user management across enterprise applications. In general, stored identifiers are used to identify and authenticate users, to ensure that only legitimate users are authorised to access permissible action. Before the advent of SSI, isolated, centralised and federated methods were the means to deploy IdMs for securing online communication and exchange of data.

### A. Isolated Identity Management

This IdMs delegates a host service to manage and maintain local registry of all users possessing unique identifiers on the service. It is mostly used by business organisation to manage identities of their employees, customers, or associates, where user attributes are tightly related to their status or identity representation on the local registry. Therefore, identifying a user is a simple local process that compares the credential presented, to that stored in the host registry for the same user. The major benefit of isolated IdM is the simplicity of its implementation. However, limitations of this model includes lack of interoperability between different services, which also hinders user satisfaction since the authentication process may vary per service. Also, since each service maintain it users, there is a burden of maintaining an exponential growth of users on the system. Likewise, the users will have to maintain, memorise or reuse credentials on every new service they access, making their passwords [114].

### B. Centralised Identity Management

In this IdM, users are bonded to a single SP based on shared secrets [10]. A user authenticates to the SP by submitting their identifier along with an authentication token (such as username and password) to demonstrate their right to be recognised by this identifier. Once the user is authenticated and authorisation is established, they are able to navigate through participating services under the SP, request for services, and have access to resources without the need to explicitly engage in a further authentication process. The scope of services a user is entitled to is not confined to a service, instead, it is bounded by the SP in which it is defined.Some limitations of this model are on SPs who keep and maintain unique security and privacy regulations, which can be cumbersome to handle for users. A classic example is the wildly disparate requirements for passwords, such as the required minimum length and the use of special characters. Also, once users delete their account, they are not able to access services. However, personal information still remains with the SP. Additionally, the centralisation of users data is a honeypot for attackers.

### C. Federated Identity Management

Federated IdM is an association for delegating responsibilities (such as authentication, authorisation, profile management, and storage services), through trust relationships between SPs and IdPs [10]. A user who authenticates to a single SP is considered to have been identified and authenticated with all other SPs, by passing assertions between SPs, based on trust established and adherence to common policies [39]. Federated IdM facilitates single sign on (SSO) using protocols (OpenID, SAML and OAuth, e.t.c), social network logins to acquire and use assertions. It also enables IdPs to keep in touch with users and offer them additional services while collecting fees from the SP they support. Additionally, federated IdM provides simplicity of management and access control in a federation.Some drawbacks to the model include the dependence on IdPs, which are large honeypots for cybercrime. Furthermore, in an event where an IdP is unable to provide authentication service, all SPs depending on it are inaccessible. Additionally, no single IdP is compatible with all SP or is able to maintain all users' attributes. Therefore, users must have accounts with several IdPs, and soon they begin to lose track of which IDP they used with whatever SP. Lastly, users' attributes are not portable across different IdPs. Public key infrastructure (PKI) based on standards like X.509 and asymmetric cryptosystems (public and private key pair) are used to secure communication in traditional IdMs. PKI's four fundamental building blocks includes digital certificates, certificate authorities (CA), registration authorities (RA) and users. A digital certificate is a means of identification used to securely authenticate users with their eIDs, they are issued by CAs after users are vouched for by RA [59].PKI is a framework that allows services to achieve a high level of information confidentiality using digital certificates. These certificates are also used for identity authentication, digital signatures and data encryption. The key pair in PKI are statistically unique and mathematically related, so a message encrypted with a public key can only be decrypted with the corresponding private key.CA and RA are responsible for the issuance, distribution, revocation, and processing of digital certificates (containing user's personal data including key pair), which raises concerns about the possible occurrence of data breaches. Other limitations of PKI includes: (1) Obtaining a digital certificate from a trusted CA might be a time-consuming operation for numerous users, especially if a local RA does not exist. (2) All communicating partners are forced to trust each other, without which communication will not take place. (3) If any trusted party fails, a process the is truncated. (4) CA centralisation is a single point of failure (SPOF). (5) Despite being a person's identity, a digital certificate can only be revoked by the CA. (6) Digital certificates only allows for the full disclosure of certificate information but not selective disclosure.

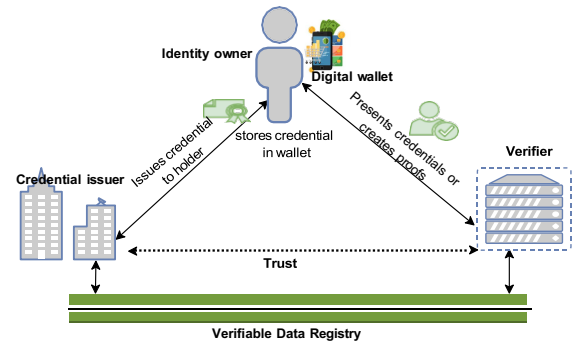

Fig. 1: SSI trust triangle, where locus of control is with the user [100]

### D. Self Sovereign identity

As the most recent development in IdMs [100], SSI enhances data privacy and integrity of traditional IdMs by enabling a holder to be free from depending on RA and CA for the issuance of digital certificates. SSI establishes a direct end-to-end peer connection between a holder and the verifying party. The connection is like a string that both holder and verifier needs to hold-on to remain active. Holders control their identifiers, giving them the ability to cryptographically create VCs from the identifiers under their control, which are used to reliably assert their identity in a decentralised way. Using zero knowledge proofs and minimal information disclosure, SSI addresses full certificate disclosure by prevent attackers from collecting private information and impersonate holders.

For example, some services only require users to provide basic information such as date of birth, email, or home address for identification, in SSI, a holder can assert this with a binary response. SSI prevents the wide spread of personal data, hence reduces the risk of Identity theft, fraud, and data correlation. It improves data privacy by putting the holder at locus of controlling the release of their data without having to go through any intermediary. Also, a verifier can confirm that identity belongs to and is linked to the owner who presents it by consulting a decentralised register. A mutual trust is maintained between the actors; the verifier trusts the issuer, and the holder trusts the issuer and the verifier. As shown in **Figure** 1, a credential issuer presents a VC to the holder, and the verifier is not required to manage or control the holder's credential information.

*1) SSI Building Blocks:* SSI is built on fundamental ideas of IdM, cryptography, distributed systems, and blockchain. The building blocks of SSI consist of seven components: actors, VC, digital wallet, digital agents, DID, blockchain, and governance frameworks [100].

*a) Actors:* SSI actors includes issuer, holder and verifier, who forms the core players involved in the creation and exchange of VCs. They make up the trust triangle as shown in **Figure** 1, and constitute the process for generating, managing, control and proof of claim of VCs. **(1)** An issuer can be a public (passport office, social security or tax office) or private (bank, university, registered business NGO) organisation that asserts information about a holder to which a VC is issued. A person can as well be an issuer (like generating and issuing a complementary card with personal details) or a thing that issues a digitally-signed credential for sensor reading. **(2)** Holder is the owner of digital identity that such as as VCs that are either self-generated or acquired from a verified issuer. It stores and control the VCs in a personal digital wallet, and presents them for proof of claim to a verifier when needed. **(3)** While a verifier, is an SP that provides services to only verified holders who meets its access policy requirements through one or more VCs that a holder presents. In SSI, trust relationship is built within these actors, giving a verifier the ability to confirm a holder's public key or DID on the blockchain or distributed registries.

*b) Verifiable credentials:* These are the electronic equivalence of our physical documents such as national cards, diplomas, work permits, drivers license and others. As defined in [121], VCs are used to claim a set of properties owned by a holder, by providing attributes acquired by the holder (such as DID, name, age, tax number), to the VC issuer, who stores them on a verifiable data registry. These attributes are verifiable from issuers for example name can be verified from the government office while tax number can be verifiable from tax office. The holder then requests the issuer to bind these attributes and generate a VC for cryptographic proof of claim (attributes, relationship, entitlement among other). To use a VC, a holder stores their attributes in a personal digital wallet [27]. A holder does not reveal their VC issuer to the verifier, just as the verifier does not learn about the holder from the issuer. However, with the issuers DID, a verifier can lookup the authenticity of a DID or VC [42].

*c) Decentralised Identifiers (DID):* These are permanent, decentralised, verifiable and unique identifiers that are used to digitally identify their owners, and a link between the decentralised owner and issuer [32]. A VC issuer attaches their public-DID to the VC they issue and stores a mirror of the DID on an immutable register (blockchain). On the blockchain, any verifier can confirm the legitimacy of the VC without having to get in touch with the issuer, even if the issuer is inactive or no longer exists. DID aims to replace single identifiers such as tax, license, passport and number among others. Typically DID is a string separated into three by two columns, the first DID prefix is a mandatory URI scheme identifier, second column is the DID identifier method, which represents the DID rules (such as Sovirn, Bitcoin, ethereum), while the last column is the DID method specific identifier. An example Ethereum DID method comprises the following syntax "did:ethr:0xb9c5714089478a327f09197987f16f9e5d936e8a"[119][37]. DID works like an IP address despite being used to identify a digital identity, it does not reveal any information about the owner after being used just once and cannot be used to connect the owner to any CA.

*d) Digital Wallet:* These are similar to the physical wallets that we store our cards and other documents in. Digital wallets are electronic portable repository applications that can store VCs and keys, they are used for secure authentication and authorisation of a holder. A holder combines attributes of one or more VCs on their digital wallet for secure presentations to verifiers [27]. Secured verification can be built on asymmetric encryption, which allows a holder to build confidence with the issuer by exchanging a key pair (private and public) [129], and the verifier by exchanging only the public key. When sending a claim, the digital wallet encrypts the claim with the public key so that a validator may decode it with the private key. For privacy and information integrity, each connection and communication holds an independent keypair.

*e) Digital Agents:* Just like we flash our physical cards or passports for physical identification to a fellow human who can read the contents of the document, digital agent oversees the process of reading through content of holders VC in SSI. There are two types of agents depending on where they are located; cloud agents and edge agents. Edge agents operate on an identity holder's local device at the network edge, while cloud agents are hosted on the cloud by verifiers such as cloud computing providers. The digital agent receives the holder's requests for private communication and employs a decentralised encrypted messaging system to enable the two different types of digital agents to exchange credentials in private communication [100]. Digital agents also provides a storage medium known as secure data stores (SDS) for digital wallet, where a holder can encrypt personal data such as files and images, which are stored and synchronised by cloud agents on behalf of the holder. SDS and digital wallets can serve as the basis for digital life management applications to process data of all types throughout a person's life and beyond.

*f) Governance frameworks:* These constitute the authorities that define policies and procedures that govern the activities of SSI actors in the processing of digital identities. They define policies such as GDPR, CCPA, and PIS. For example on a regional level, the European Union (EU) member states, in collaboration with the EU commission, launched the Europe Self-Sovereign Identity Framework (ESSIF), with the goal of establishing a GDPR-compliant framework to assist the European digital single market.

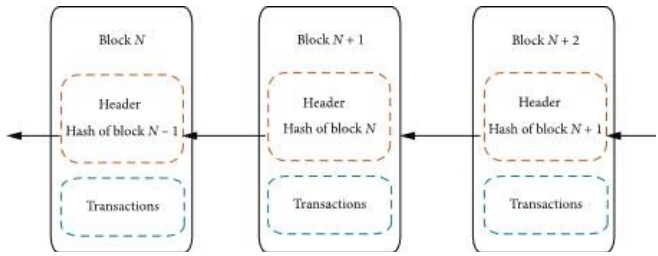

Fig. 2: A simple description of blocks of transactions in a blockchain network [116]

*g) Blockchain:* In 2009, blockchain attracted the world's attention through the first known real-world application of bitcoin cryptocurrency [86]. Blockchain is a distributed system consisting of immutable public databases, that are cryptographically hashed in peer-to-peer public transactions that are not managed by a single centralised entity, instead records are publicly available and trustworthy to all network users. Traditionally, blockchains are classified into two: public and private [116], as well as permissioned and permissionless. A public blockchain is entirely open to all public members, whereas a private blockchain is only accessible to approved entities. Since it's publicly accessible, blockchain rarely uploads PII (personal identifiable information) to the network. In a permissioned blockchain, all entities in the network have read access, but only a subset of the entities have written permission. A permissionless blockchain allows anybody to write to the blockchain. Ethereum and Bitcoin are instances of public permissionless blockchains since they can be viewed and written by anybody. In bitcoin users can add a personal transaction at the end of the blockchain in a consensus manner, and other users can read the transaction list to learn who owns which bitcoins [77]. Ceramic network [25] and Hyperledger indy are examples of public permissioned blockchains [13]. Blockchain shares a consensus (such as Proof of Stake PoS, Proof-of-Work, PoW, Proof-of-Authority PoA, Practical Byzantine Fault Tolerance PBFT, and Delegated Proof-of-Stake DPoS) algorithm that allows transactions to be completed, synchronised, and generates an ordered list of stored and associated information through a chain of blocks that usually contain the previous hash block, data content, the participant signature, and the timestamp. The previous hash block causes the information in the blockchain to remain immutable. Blockchain data may include payment information (such as Bitcoin and Litecoin), contracts (such as Ethereum), or personal information, and many more. This is described in **Figure** 2.

## III. RELATED WORKS

Blockchain technology and SSI frameworks complement each other, making a perfect symbiosis. Blockchain provides the necessary method of decentralisation with an immutable registry of information required for SSI frameworks, which are used to implement smart contracts, store identifiers, proof of ownership, binary relationships (guardianship and delegation), certificate revocation list (CRL) and serve as a source of cryptographic verification, among others. As a new technology some survey works have reviewed SSI based solutions that aims to address data breaches. Herein we provide an analysis of some review works and compared them with our work in **Table** I.

Karasek et al in [60] conducted a study to reconcile the methods employed in permissionless distributed ledger technology (DLT) which complies with GDPR and other financial rules such as the anti-money laundering (AML) and combat terrorist financing (CFT). The authors suggested a new policy in which decision-makers would work together to choose and put AML/CFT measures into practice. Governments and virtual asset service providers (VASP) would be able to avoid creating privacy-focused networks that employ technologies and enhance anonymity, like zero-knowledge proof and cryptography.

Haque et al in [50] conducted a systematic review on GDPR compliant blockchain systems. The authors used 39 research articles which they divided into six categories according to the GDPR article they sought to address. The categories are; data deletion and modification, data protection by design, responsibilities of controllers and processors, consent management, data processing, principles of lawfulness and territorial scope. They further divided these publications into subcategories based on the subject areas they were studying, which included smart cities, information governance, human identities, financial data, IoT, healthcare, and online data. Based on the their analysis, the authors found that GDPR's article 16 and 17 on data deletion and modification are among the most researched compliance topics for blockchain.

The authors in [4], acknowledged the advances in data security where entities can adopt a privacy-friendly technique for managing customers' data by switching from centralised data repositories to a trusted decentralised method using blockchain. However, they argued that GDPR focuses on centralised data management, making blockchain not suitable. The authors analysed blockchain solutions that aim to enhance data privacy, they then addressed limitations of the studies by proposing a blockchain method that complies with GDPR's policy on data protection.

Freund et al in [44] examines how data is handled in blockchain technology according to the phases of the data lifecycle (DLC), which they categorised as; collect, storage recovery and disposal. The authors studied 7 articles and evaluated their compliance level with GDPR data privacy principles. Particularly, how data relates to or depend on one another during various stages of a process in the categories. Based on the limitations from their evaluation, the work presented a technology to adopt DLC which is guided by GDPR privacy principles.

Politou et al in [97] noted how works incorporating blockchain, which has immutability, are in conflict with the GDPR's right to be forgotten. They provided a comprehensive review of solutions in the literature that attempted to address this contradiction through technical methods and advanced cryptographic techniques. The study also discusses the potentials, restrictions, and limitations of the solutions when used in the real world with either permissionless or permissioned blockchains.

| | Systematic review | GDPR privacy-principles achieved | Covered right of data subject | Compared review works | Blockchain enabled | Covered data delegation solutions |
|---|---|---|---|---|---|---|
| [60] | **No** | Six | Yes | **No** | Yes | No |
| [50] | Yes | One | Yes | **No** | Yes | No |
| [4] | Yes | Four | Yes | **No** | Yes | No |
| [44] | **No** | Yes | **No** | **No** | Yes | No |
| [97] | No | None | Yes | **No** | Yes | No |
| [8] | No | one | Yes | **No** | Yes | No |
| [16] | Yes | None | No | No | Yes | No |
| [64] | No | None | Yes | **No** | Yes | No |
| [3] | No | None | No | Yes | Yes | No |
| [38] | Yes | Yes | Yes | **No** | Yes | No |
| [110] | Yes | None | Yes | Yes | Yes | No |
| This work | **Yes** | **Yes** | **Yes** | **Yes** | **Yes** | **Yes** |

TABLE I: Comparison of some review works in the literature

Ahmed et al in [8] examine the compliance of online social networks (OSNs) particularly Facebook with GDPR's principle of seeking data subjects' explicit consent for different purposes. The study analysed methods adopted in the literature and identified vulnerabilities that could potentially result in GDPR non-compliance. Although GDPR advocates for informed consent to process personal data, the authors argue that a good number of works in OSNs do not meet the informed consent criteria for GDPR, neither are the methods adopted fine-grained enough for users to understand.

Solina et al in [16], provided an overview of the literature on the use of blockchain-based systems in transportation. The authors goal was to identify current research trends, major gaps in the literature, and potential future challenges using a multi-step approach. The authors retrieved pertinent works from the literature that they considered to be the most important in transportation fields for: Supply chain management and logistics, road traffic management, and smart cities. They analysed these documents and observed that although blockchain technology is still in its early stage, it already shows great promise in a variety of fields, including supply-demand matching, food track-and-trace, regulatory compliance, and smart vehicle security. Additionally, the relationship between blockchain and sustainability was explored, and they demonstrated how this technology might lead to reduction in food waste, exhaust gas emissions, promote proper urban development, and generally enhance quality of life. The authors also noted that only a small proportion of blockchain proposals have been implemented in practise.

Galia et al in [64] explored three blockchain technical solutions: Sovrin, Hyperledger Indy, and Jolocom. They examined the method these blockchains uses in SSI solutions, as well as their adherence to GDPRs rules on data controller, credential revocation, DID management, justification of usage, and right to be forgotten. The authors came to the conclusion that as SSI involves the processing of personal data, a thorough investigation of the solutions and their use cases is necessary to determine compliance with GDPR.

Adam et al in [3] reviewed methods for traditional and decentralised access control in IoTs. Although traditional access control systems relies on CAs, it supports access control and authentication procedures with ease. The authors pointed out that because CAs are vulnerable to SPOF and other security and privacy issues, they do not scale well for the future growth of the IoTs. The decentralisation feature of blockchain addressed many of the issues raised. The authors also pointed out certain unresolved problems with this method, including privacy leakage, the limited resources of IoTs and inadequacy of access control enforcement. Driss et al in [38] conducted an extensive study of privacy-preserving solutions within the smart healthcare domain. The proposed solutions' architecture, privacy principles, and stakeholder needs, among other things, were examined. From the analysis the authors came to the conclusion that out of 100 primary studies, more than 70% had neglected the needs of the stakeholders, particularly the adherence to patient privacy preferences and privacy laws, as well as the non-compliance with GDPR's privacy principle of data collection limitation and data minimisation.

Schard et al in [110] conducted a systematic review of new SSI challenges, categorising them as either pragmatic or conceptual. Among the pragmatic issues are methods for migrating traditional IdMs to SSI, trusting data from sovereign identities, and managing digital credential recovery and backup. The conceptual issues captures how SSI systems are defined and what constitutes them. To address these concerns, the authors posed some research questions, gathered and analysed 82 articles. Based on the issues, they proposed a taxonomy and systematic mapping of the solutions. They further classified related articles introducing and solving novel problems in the SSI ecosystem into four categories: management, operations, system design, and trust. They also categorised articles that introduce mathematical approaches to SSI issues, such as cryptographic and non-cryptographic tools.

## IV. RESEARCH METHODOLOGY

Secondary studies are needed to keep-up with changes in main research efforts on any particular topic. Therefore, using the systematic review guidelines discussed in [61], we propose some research questions and sourced for relevant literature on the questions. Retrieved documents were screened using the inclusion and exclusion criteria pre-specified as shown in **Figure** 3.

Our main research questions (RQ) include:

1) How has the evolution to SSI and use of blockchain-based systems complied with GDPR guidelines in the issuance, ownership and processing of VCs?
2) Are SSI and blockchain solutions consistent with GDPR requirements in managing user consent, privacy control and certificate management?
3) What are the literature approach to revoking VCs, and to what extent do they comply with GDPR principles?
4) What are the current literature approaches that utilises SSI-based technology for delegation?

*A. Data sources, inclusion and exclusion criteria*

To address the outlined questions, a systematic search for relevant English-language articles that are digitally accessible in full on databases such as Web of Science, Scopus and IEEE Xplore. Our search was limited to these databases, as they publish peer-reviewed article (journals, magazines or conference proceedings) and permit logical expressions (keywords, names, and/or abstracts). Additionally, the articles must provide a case study or offer a framework on GDPR compliance challenges with SSI or blockchain. We removed duplicate articles, review and thesis works, and articles that did not discuss SSI, GDPR, or information privacy and security issues.

*B. Search Strategy*

The concept of SSI is less than a decade old and is still evolving. Our search captured the most recent works, our search captured electronic literature published between January 2019-to-June-2021, our search was restricted to the tittle, abstract and keywords of the papers. We conducted the search using the following keywords :("blockchain AND GDPR AND data AND processing"), ("blockchain AND erasing of certificates"), ("blockchain-based certificate AND revocation") and ("delegation with blockchain").

*C. Screening and Final Selection.*

From the search conducted, 402 documents were retrieved. We filtered out the duplicates and redundant documents. To ensure that retrieved documents are within the scope of our study, we subjected the documents to an initial screening process where we skimmed through the tittles and abstract, after which we filtered out non-relevant papers, leaving us with 143 documents. From these we selected primary research documents and read through the abstract, introduction and conclusion, based on their relevance we further excluded 95 documents. Again, we analysed contents of the remaining documents and filtered out survey papers and documents with no defined research questions. Finally we selected 48 articles that are discussed in this work as shown in **Figure** 3.

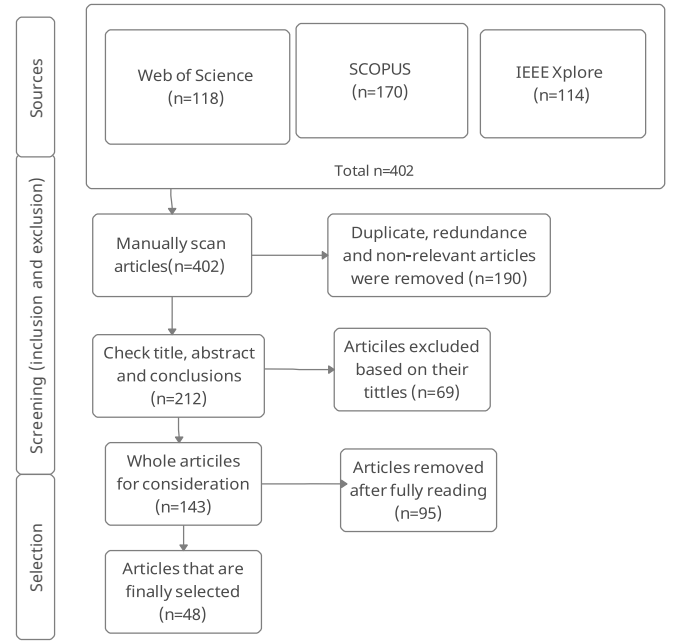

Fig. 3: Summary of articles retrieval and screening process

V. THE RIGHTS OF DATA SUBJECTS UNDER GDPR

GDPR came into effect on 25$^{th}$ May 2018 as the major update to the European data protection law which has been in use for over 20 years. It focuses on protecting the right of data subject's privacy and the use of private data which include data creation, receipt, transfer, ratification, deletion and delegation. With the enactment of GDPR, the need to protect European users data became a legal requirement for all data processors within and outside the EU's digital space [96]. GDPR includes seven core principles outlined in article 5 of the regulation (transparency and awareness, purpose limitation, accuracy, control, auditability, data integrity and data minimisation). For this purpose, it forwards 26 definitions including; (1) personal data, as any information relating to an identified or identifiable natural person ('data subject'), which is processed by controllers, (2) data controller, a digital identity that, alone or in collaboration with others, sets the goals and means of personal data processing, (3) data processing, as any operation or set of operations performed on personal data from receipt to storage, dissemination, erasure or destruction, (4) restriction, tagging of stored personal data intending to restrict its future processing. Observing these requirements while processing personal data is mostly case-specific. For example, GDPR encourages the use of data pseudonymisation and anonymisation to prevent data correlation. However, both serve different purposes in specific scenarios [100]. In spite of the numerous benefits provided by SSI and the importance of regulations such as GDPR, CCPA and PIS, there has been no single globally acceptable standard or framework to evaluate SSI and blockchain compliance with these regulations [100].

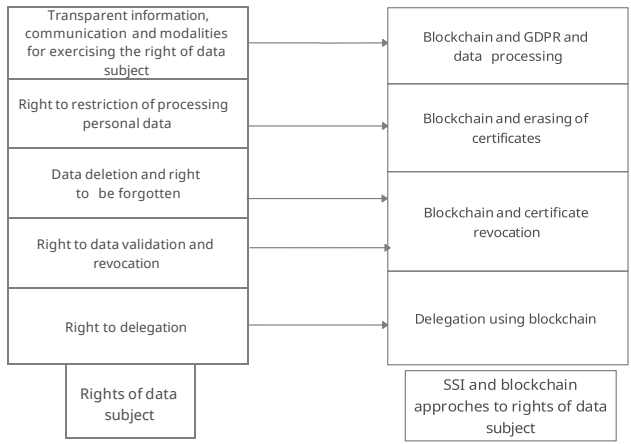

Fig. 4: Rights of data subject and approaches by related works

### A. GDPR methods To Protect The Use Of Personal Data (Right of Data Subject)

Due to the limitations of contextual enhancement of systems by users, the non-adherence of data controllers processes to GDPR policies can have a negative impact on adoption of systems by users, since they would have a limit to which they will be able to comply with GDPR. This is especially important when processing high-risk data like health and financial records [66][1].

In order to achieve this compliance, data controllers are mandated to ensure that a preliminary design of systems involved in processing people's data must adhere to GDPR policies in all processes. For a data breach, GDPR directs that data subjects should be notified in not more than 72 hours of such occurrence. GDPR comprises several chapters containing different aspect of data processing. However, in this document, we concentrate on 5 aspect of data subject's right to address our research questions as illustrated in **Figure** 4.

### B. Addressing RQ1

Our first search query, "blockchain AND GDPR AND data AND processing" is considered for this research question. To address this, we utilised GDPRs "transparent information, communication, and modalities for exercising the rights of data subject". We outline related literature for this section in **Figure** 5 and summarise in **Table** II.

*1) Transparent information, communication, and modalities for exercising the rights of data subject:* In GDPR, the roles of controller and data subject are clearly defined. However, entities assuming these roles are use-case specific and thus need different capabilities [75]. For example, controller's role in health, banking and educational institution varies due to different sensitivity of data used. The role of data controller also differs between on-chain and off-chain data storage. In off-chain solutions where personal data does not reside on a blockchain, data subjects could assume the role of controller. However, in on-chain solution where data reside on the blockchain and is interoperable, many entities could assume the role of joint controllers including data subjects.

In either case, GDPR's articles 6, 12, and 15 outline the prerequisite and justification needed to process personal data which includes: receipt, access, processing, dissemination, storage, and erasing of personal data that have to be identified and consented to by the data subject or representative as prescribed within the legislation. Additionally, controllers are to build systems with privacy-by-design, through technical and organisational measures to ensure that only personal data necessary for a particular procedure are processed. Although GDPR provides sufficient legislation to protect data subject's consent and processes of controllers, the justification to whether these consents decision are adhered to in processes is mostly beyond data subject's purview. Despite the GDPR's implementation, compliance has been a serious challenge since most services lack the necessary infrastructure to enforce the essential procedures. As a result, the authors in [111] suggested a framework to assess the compliance of services by intercepting data requests using a GDPR-compliant blockchain system that securely monitors data activities while protecting owners' privacy. Authors in [66], also proposed a federated blockchain network with a smart contract that enforces GDPR requirements. The network keeps the traceability of health data shared between interconnected institutions (health, research, insurance among others) throughout the data life cycle. To provide a transparent platform for data owners' awareness to right on their personal data, and record the consent they have provided, the authors in [30], proposed a lightweight blockchain-based GDPR data management system, granting access to immutable evidence of agreement between data owners and SP. Where data subjects are informed of what happens to their personal data and are able to regulate it in accordance with their consent, and SP can show that they are complying with the regulation.

Using a combination of private and public blockchains, the authors in [109] proposed to address transparency and privacy issues in processing personal data that relates to machine learning and artificial intelligence drive for automated decision making. The public blockchain is designed to serve as a trust anchor for the private blockchain, whereas the private blockchain stores log of access control of personal data, which is under the control of the owner. In [99], the authors integrated attribute-based encryption with private blockchain technology to propose an architecture for sharing and storing medical data. It also uses a smart contract to address confidentiality, storage and access control issues. Data store on the private blockchain is completely invisible, and access to it needs a signed smart contract that must be produced by the user.

Awuson in [19], acknowledged the challenges faced by cloud forensic investigators handling criminal cases in acquiring digital evidence from online services. To process and validate all log transactions, they proposed a permissioned blockchain cloud forensic login framework which maintains data immutability and encrypts transaction that are hashed.

The work in [118], aimed to address the centralisation of data from different sources used in traditional machine learning environments by introducing a federated learning system which they termed "DFedForest" (decentralised federated forest). The system is distributed, as it generates random forests across domains in a federated way, and uses permissioned blockchain to register mutual trust of data sources and reference local model addresses in a distributed way that prevents malicious participation that might harm the accuracy of the model. Currently, online services provide non-granular, hard to read, and long agreements requiring users' consent before services are granted [88]. Sometimes, these agreements are written in ambiguous terms, making it unclear as to what the agreements mean or whether a particular firm actually agrees to it. Despite the ambiguity, users consent to such agreements without going through or fully understanding the consequences. To address these issues, the authors in [45] summarised data subject's rights in GDPR and CCPA into three; (1) right to consent, rectify, and delete, (2) right to be informed, and (3) right to access and transfer. They then proposed a provably compliant user rights manager using a two-layer blockchain. The system defines users right as formal executable sharing agreements that can be automatically translated into a human-readable form and enforced when data is requested. An indelible, verified trail of the hash of data access and sharing agreements is kept on a two-layer distributed ledger to support revocation and show compliance. The main chain guarantees partition tolerance and availability (PA), while the side chains assure consistency and availability (CA), resulting in the CAP's three characteristics (consistency, availability, and partition tolerance).

| Literature description | Year | Domain | Technology | Contribution |
|---|---|---|---|---|
| Enabling trust in healthcare data exchange with a federated blockchain-based architecture [66] | 2019 | eGovernment | Blockchain and spring framework | A framework to assess the compliance of services to GDPR by intercepting data request |
| Protecting and assessing of GDPR compliance on private data [111] | 2020 | Healthcare | Federated blockchain, catalogues and CAs | Trust in healthcare data exchange |
| Lightweight Blockchain-based Platform for GDPR-Compliant Personal Data Management [30] | 2021 | Generic | Lightweight Blockchain | Auditing and transparency towards data owner and privacy incident/breach detection |
| Transparency with respect to personal data handling [109] | 2019 | Generic | Hybrid blockchain (private and public) | Pubic access to immutable agreement between data owner and service provider |
| MedSBA: a novel and secure scheme to share medical data based on blockchain technology and attribute-based encryption [99] | 2020 | Healthcare | Public and private blockchain ciphertext-policy attribute-based Encryption and ABAC | Secure sharing of medical data with enhanced privacy consideration |
| BCFL logging: An approach to aquire and preserve admissible digital forensics evidence in cloud ecosystem [19] | 2021 | Forensic investigation | Blockchain Cloud Forensic Logging (BCFL) framework and smart contract | Permissioned blockchain to maintain tamper-proof log evidence in the cloud ecosystem |
| Distributed machine learning framework with local random forest algorithms and decision trees through the blockchain. [118] | 2020 | Generic | Random forest algorithm and hyperledger fabric | A distributed random forest and federated method that avoid sensitive data leakage |
| Soteria: user right management framework to protect private data with enhanced privacy [45] | 2020 | Generic | Distributed ledger and Audit trail | Executable sharing agreements that specifies users rights under GDPR and CCPA |

TABLE II: Literature works addressing RQ1

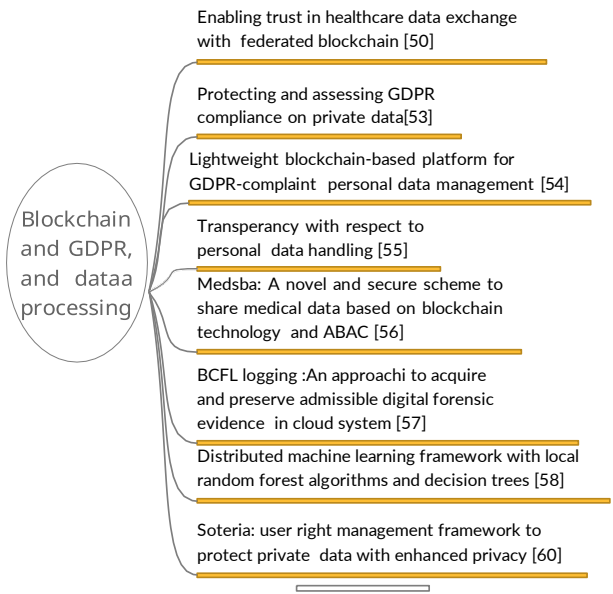

Fig. 5: An overview of proposed solutions for data processing with blockchain and GDPR

*2) Existing practical challenges:* Works in this literature sought to provide solutions to the processing of personal data which include data privacy, integrity, transmission, and transparency. However, the use of blockchain registers poses a contradiction to GDPRs' requirement of requesting data owner's consent for processing personal data. Commitment sometimes also affects the user's right to explicitly issue consent that is free from any control or monitoring data use. These are open reasons for future research. Key open research issues include.

1) Transparency in data processing: Blockchain is a network of distributed databases where miners can get a copy of the chain and possibly add records without changing existing data provided the chain is active. This implies that active members of an active chain can process data on the chain belonging to others without the owner's consent. This contradicts the data owner's right to transparency in processing personal data.
2) Data privacy and anonymity: Despite being used for identifying a holder, DID and other cryptographic commitments prevents out-rightly providing any information on the owner, nor can the owner be traced to any central authority [32]. Therefore, when used only once, there is a limit to the information that can be released to a service. However, when used multiple times, the DID works like an identifier, it allows linking of different credentials or attributes which if correlated, can identify a holder [65]. This creates a need for methods that prevent the reconstruction of DIDs from distributed storage.

3) Storage centralisation: The work in [118] [66] uses domains and central endpoints that federates data across each other. However, once a domain is inactive, the information it emits remain inaccessible. In such scenario, the use of a distributed file storing such as IPFS with integrity checks would be useful.

### C. Addressing RQ2

To address this RQ the second search query, "blockchain AND erasing of certificates" is taken into consideration. In accordance with GDPR requirements for the "right to restriction of processing personal data", the literature was organised into three which are: (1) secured data correlation, (2) data consent, privacy and integrity, and (3) data deletion and right to be forgotten. GDPR provides for right to data privacy [96]. Categorically, data subjects should have supreme control over their data, and be free from any third party, be it an individual or organisation. This should cover any means that require the usage and correlation of their data across domains and should only last for as long as they desire to be known. Furthermore, data subjects should have the right to have inaccurate personal data corrected or rectified. However, this should not contradict the user's right to delete their data (right to be forgotten) as mentioned in GDPR's article 17. Also, these processes should not affect the right to restriction that requires permission from data subject. We discuss the literature retrieved in the sections below and outline them in **Figure** 6 and **Table** III.

*1) Secured data correlation:* Since the inception of GDPR, the right to privacy while processing personal data became a legal right. Pseudonymisation is one way of concealing the meaning of private data, while data minimisation ensures that data collected about an entity is only used for the purpose of collection and nothing more. These methods and cryptographic commitment are used to secure proof of identity. Additionally, users can commit their choice to a process or transaction (by issuing part of their credential) without making such choices public or revealing their identity. Secret sharing, zero-knowledge proof, secure signature and others are examples of cryptographic commitment [29].

In SSI, the creation of DID and VC replaces single identifiers such as tax numbers, license numbers, passport numbers and others. Typically, DID is free from any central authority for creation, processing, and storage but depends on a decentralised registry (such as blockchain and DID methods) for a cryptographically verifiable identity [100]. It promotes the use of anonymity, pseudonymity, and different identifiers for proof of identity, making it significantly more difficult to correlate the holder's activities on services. It also supports selective disclosure to enhance risk decentralisation and trust-less service, where proof of identity is divided and secretly shared among verifiers, from which only qualified verifiers are able to reconstruct the required identity attributes for the proof. Zero knowledge proof is another cryptographic method used in SSI to control unwarranted access to private data. It allows a data owner to proof the ownership of specific identity attribute such as age of maturity, membership affiliation, location and others without revealing the exact certificate itself. This enhances data owners privacy, data integrity and prevents data correlation as verifier does not learn beyond what is provided, neither are there links to the credential issuing process. Equally, secure signature is used to protect the identity of a signer to a secret message by signing the message with a combination of public key and unspecified amount of private key.

To protect personal data in shared resources, a consolidated architecture is proposed in [94]. The architecture combines key management protocols for secret sharing of personal data, pseudonymity and data minimisation are further used for identity masking, where the same identity reveals different information to different services about a person (to avoid linkability), digital signatures are then used to protect users profiles. In [113], the authors proposed a framework for India's national healthcare service, that interoperates health data between institutions for transparent insurance claim using a smart contract. The system allows patients to authenticate using zero-knowledge proof and delegate claim services using proxy re-encryption to a cloud service. Authors in [107] acknowledged the privacy issues and constrains with storage for IoT devices. They proposed an effective authentication system with off-chain storage to ensure user privacy, hence complying with GDPR. The solution is a blockchain-based IdM for the smart industry, which relies on assured identities from anonymous biometric credentials to ensure secured data correlation and selective disclosure of credentials. They did this by utilising suspension of credentials, and their permanent revocation which are added to a credential revocation list (CRL) on a blockchain, while merkle tree roots are used to audit the CRLs. Also, the authors in [51] proposed a system for GDPR compliant health data exchange that relies on two blockchain design, one with an IOTA; a public distributed ledger, while the other adds a private IPFS cluster to the ledger. In their view, using the IPFS minimises the risk of data misappropriation, with satisfactory performance for the exchange of larger health records specifically blood glucose data.

*2) Data consent, privacy and integrity:* In [31] the authors proposed a blockchain based model for GDPR consent compliance verification. Their method is divided into four components: (1) they used semantic ontology to formalise GDPR consents issued by data owner, (2) a decentralised personal data management based on hyperledger fabric is used to store compliance to data owner's consent using smart contracts, (3) a data repository is built with MangoDB to enforce the legal privacy requirements of consents, and (4) they used XACML to confirm compliance on personal data access control methods. On-chain data created in [5] are encrypted with private keys of the data owner, giving them full control of their certificate in an SSI fashioned manner. The system also empowers data owner with selective disclosure of personal data. Using smart contract and a permissioned ledger, a similar technique is proposed in [120]. The authors in [76] proposed a blockchain-based innovative framework to protect data privacy, provide availability and integrity of data processed in IoTs (smart-cities, healthcare and others). In their proposal blockchain network is divided into different channels that contain data used by a specific organisation for example data marked for health can only be used by health institutions, while data marked for education can only be used for academic and research purposes and so on. Access to data on each channel is then protected by embedding access control rules in a smart contract. The authors in [12] aimed to achieve a decentralised SSI framework while addressing data security, privacy, and performance issues of Sovrin[119], Uport[52] and Shocard[74]. In the framework, the authors proposed a privacy-preserving architecture using blockchain-based approach. They utilised the domain name system (DNS) experience and smart contracts to enable data owners to maintain certificates that are associated with certain attributes. Their proposal eliminates data centralisation and linkability from third-party, which allows SP to identify data owners and provide personalised service while retaining the user's ownership of their data. In [125], the authors proposed a blockchain-based data privacy management framework that aimed to address the breach in managing customer data within open banking services. The framework is combined with nudge theory, it further dissolves the default method of data disclosure scheme of new customers with a collaborative filtering method that receives and processes customer data based on the segregation.

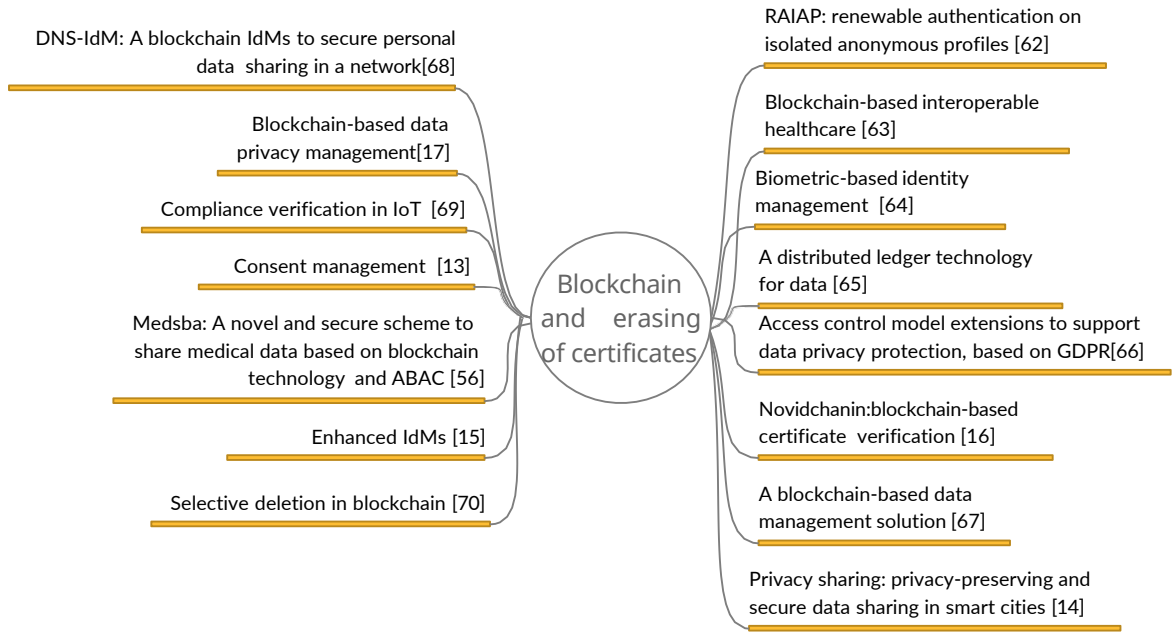

Fig. 6: Featured works on erasing certificates with blockchains

In [21], a data privacy framework aimed at improving the visibility of private data amassed by IoT devices is proposed. The solution adopted the use of smart contracts as blockchain transactions to verify the compliance of IoT devices with GDPR, while collating data before been accessed or manipulated. Ontology mapping, fine-grained access control, and IoT devices are also used in [102] to implement data access accuracy, storage limitation, and provide data integrity. The authors in [99] integrated attribute-based encryption with private blockchain technology to propose an architecture for sharing and storing medical data. The work also uses a smart contract to address confidentiality, storage and access control issues. Data stored on the private blockchain is completely invisible, and requires a signed smart contract presented by a user for authorisation of access to such data. In [27], a decentralised identity management framework is proposed to give data owner control of their data. The proposal has been piloted for digital staff passport in UK's national health service by extending VCs with FIDO2 authentication. The framework allows both patients and health care practitioners to establish a FIDO2 link with a VC issuer for decentralised key generation, that are stored in users portable device. The issuer stores attributes such as training skills, professional status, then assigns VCs that contain this personal attributes. For health workers, it enables the secure on-boarding and sharing of personal details when needed under their control. It also eases the access to much-needed medical records for consultation, issuance of prescriptions, appointment, or cancellation. For patients, they can book or cancel medical appointments for consultations, request for prescribed drugs and others at their comfort without the need to visit a health care center using their VC from their digital wallet. In their view the solution provides selective disclosure of attributes as both verifier and data owner are able to agree on set of attributes needed for identity proofing. It also provides attribute aggregation without requiring any complete revocation mechanism or blockchain infrastructure.

| Proposal description | Year | Domain | Technology | Contribution |
|---|---|---|---|---|
| RAIAP: renewable authentication on isolated anonymous profiles [94] | 2020 | Generic | Pseudonymity, key-management, non-repudiation and data minimisation | A theoretical framework established on ideas from SS-IDs and DLT |
| Blockchain-based Interoperable Healthcare [113] | 2020 | Healthcare | Inter Planetary File System (IPFS), smart contracts and Zero-Knowledge Succinct Non-Interactive arguments of knowledge (ZkSnarks) | Smart card for authentication using ZkSnarks and delegate access to service the providers via proxy re-encryption |
| Biometric-based identity management [107] | 2021 | Smart industry | Fuzzy extractors, offchain storage and accumulator for improved scalability | Biometric-based anonymous credential scheme that supports selective disclosure, suspension/thaw and revocation of credentials/entities |
| A distributed ledger technology for data [51] | 2019 | Healthcare | IOTA distributed ledger and IPFS cluster | An immutable, interoperable, and GDPR-compliant exchange of blood glucose data |
| Access control model extensions to support data privacy protection based on GDPR [31] | 2019 | Generic | Provenance Ontology, Hyperledger Fabric, MangoDB and XACML | Blockchain-based model for compliance verification under GDPR |
| NovidChain: Blockchain-based certificates verification [5] | 2020 | Healthcare | Ethereum blockchain, DIDs and IPFS | Blockchain-based privacy-preserving issuance and verification of COVID-19 test/vaccine certificates |
| A blockchain-based data management solution [120] | 2019 | Generic | Hyperledger fabric and public-key hierarchical model | GDPR-compliant blockchain-based personal data management |
| Privysharing:privacy-preserving and secure data sharing in smart cities [76] | 2019 | Smart cities | Blockchain network with dual security in the form of an API Key and OAuth 2.0. | Blockchain-based secure and privacy-preserving data sharing mechanism for smart cities |
| DNS-IdM: A blockchain identity management system to secure personal data sharing in a network [12] | 2019 | Generic | Domain name system, Ethereum and IPFS | A decentralised IdM architecture with attribute associated validators |
| Blockchain-based data privacy management [125] | 2020 | Banking | n-chain and off-chain switching algorithms, nudge theory | Blockchain-based data privacy management with Nudge theory in open banking |
| Compliance Verification in IoTs [21] | 2020 | IoT | Smart contract | GDPR compliance verification for IoTs using smart contract |
| Consent management [102] | 2019 | IoT | Ontologies, proof of authority (PoA) algorithm, and eXtensible Access Control Markup Language (XACML) | A Blockchain-Based platform for consent management of Personal Data Processing in the IoT Ecosystem |
| MedSBA: a novel and secure scheme to share medical data based on blockchain technology and attribute-based encryption [99] | 2020 | Healthcare | Public and private blockchain ciphertext-policy attribute-based Encryption and ABAC | Secure sharing of medical data with enhanced privacy consideration |
| Enhanced identity management system [27] | 2019 | Healthcare | VC and FIDO UAF | Improved identity management with verifiable credentials and FIDO for strong authentication and authorisation |
| Selective deletion in a Blockchain [54] | 2021 | Generic | Consensus algorithm for chain deletion | Creating a summarised chain and deleting unwanted information |

TABLE III: Literature works addressing RQ2

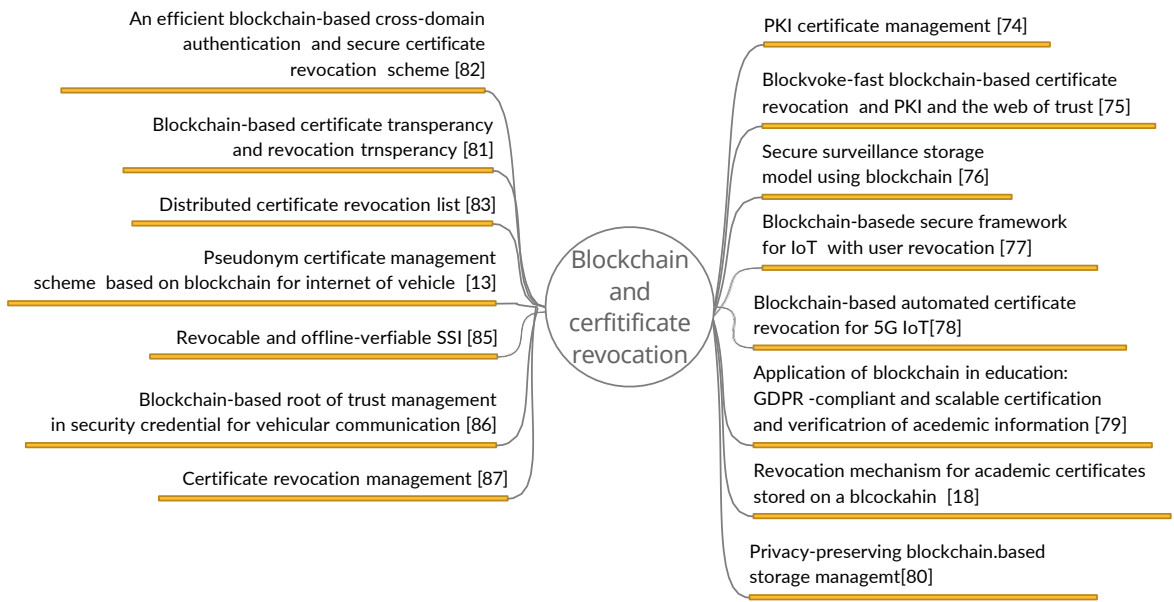

Fig. 7: An overview of solutions for blockchain and certificate revocation

*3) Data deletion and right to be forgotten:* In traditional IdMs that use identity federation with a certificate that is controlled by an IdP, exercising the right to data erasure might be straightforward since the data owner's certificate will be deleted from the IdP. In SSI, possible solutions might differ depending on the storage method. In off-chain solutions, data owners control access to their VC, since the VCs reside on personal digital wallets and can easily be erased, therefore exercising their right to be forgotten. However, a hash/signature of the VC is stored on the chain or a verifiable data registry and remains there provided the chain/registry is active. Therefore, with immutability of blockchain, deleting such record can be complex (if not impossible). Although blockchain property of data immutability is regarded as one of its benefits, it contradicts data subject's right to data deletion and right to be forgotten [77]. To address these issues, the authors in [54] proposed a selective blockchain deletion method for deleting records in a blockchain while retaining key qualities such as trustworthiness and consistency. The solution uses a consensus algorithm for chain deletion that is extended by regularly creating summary blocks of previous on-chain data, which are stored again in a new block leaving out unwanted, redundant, unused or deleted at the end of the blockchain.

*4) Existing practical challenges:* Although current literature attempts to provide users with medium to exercise data privacy and fundamental right to erase personal data, some drawbacks to this attempt exist, which include the following:

1) Data leak: Being new systems, most of which were created within a short period of time, some frameworks such as Covid-19 test and vaccine applications have been flawed with data privacy concerns of tracing citizens, [130][79], sharing personal health information, and can be prone to false positive and negative errors [9].
2) Digital identity monitoring: Solutions in [102][21] used IoT devices for collating digital identities and smart contracts to verify compliance with GDPR. However, IoTs comprise devices and sensors used in people's daily lives for many activities such as health monitoring, precision farming, transportation and others. These are vulnerable to security attacks such as man-in-the-middle, denial of service (DOS), social engineering, among others [89]. Moreso, IoT devices tend to emit other information such as time, location and others (users are mostly unaware). Sometimes these are sent to unknown third parties without owners' consent, which can independently identify an entity or if collated with other information use to trace or monitor digital identities. These contradicts right of digital identities to privacy.
3) Storage and data flexibility: The solution in [5] uses blockchain, DID method and IPFS for obtaining and preserving Covid-19 test result. A drawback of this framework is the use blockchain for storing users. Because of the immutable property of blockchain patients cannot enforce data minimisation on their health record. Also, health institutions can access patients' records against their wish even after they fully recover from the virus.
4) No-uniform data: Although the work in [27] proposes a solution to address data use, it does not address inconsistency in minimal data acceptance for identification and authorisation. There is a need for ontology classification for issuers, users, and services to define data standards.
5) Centralisation and delegation: Patient verification completely relies on private key from a centralised hospital in [113]. Likewise, the proxy re-encryption server depends on a delegated server, which both can be complicit or suffer from SPOF

### D. Addressing RQ3

The third search query, "blockchain-based certificate AND revocation" is considered for this research question. In accordance with GDPR requirements for data subjects right to data validation and revocation, we categorised the works into two: (1) revocation management, and (2) validation efficiency and distribution. **Figure** 7 and **Table** IV provides a summary of this section.

Access revocation is the act of invalidating a consent issued for representation to process personal data or transaction. GDPR article 7 provides for this right to data subject. Conventionally, a document or VC is issued over a period and naturally ceases to function once expired. However, specific situations might warrant the need to revoke a document prior to its expiration for reasons such as; VC no longer used, attributes of a VC have changed, compromise in authorisation process or any suspected fraudulent behaviour among actors. Also, a University diploma has no expiry date but can be revoked by the University for suspected fraud, misuse or forgery. Likewise, an organisation can authorise their employee and issue a VC to carry out authorised activities on their behalf, however the organisation can revoke the VC once the employee is no longer with them, but not because it has expired. Once an issuer revokes a VC, its function seizes and the data subject is denied access to services with it. Currently, there are some conventional methods to check a revoked certificate such as online certificate status protocol (OCSP) [82], certificate revocation list (CRL) [105], certificate revocation tree (CRT) [63] and CRL distribution point (CRL-DP) [80]. While these solutions mostly meet their goals as source of attesting to the validity of certificates, their implementation on SSI technology raises scalability, security, and privacy issues [7]. For example, the revocation does not mean a withdrawn certificate has been deleted, rather a centralised list of revoked certificates with attributes of both issuer and the data subject has been created. Also, centralisation of revoked certificates are prone to data breaches such as man-in-middle and DoS [90]. Moreover, the periodicity of CRL update could lead to use of a revoked certificates before CRLs are updated [71]. Also, as the list of issued and revoked certificates grow exponentially, centrally managing this could affects the effectiveness of services and possible causes credential overflow [23]. Furthermore, the traditional implementation of these methods only considers the certificate issuer for revoking VCs.

With decentralisation in SSI, revocation can be implemented by the owner since VC is not under a central entity. However, on-chain implementations stores revocation lists in blockchain

| Literature description | Year | Domain | Technology | Contribution |
|---|---|---|---|---|
| Public key infrastructure certificate management [47] | 2020 | Generic | Blockchain and PKI | A registration and certificate authority management system for certification processes |
| BlockVoke-fast, blockchain-based certificate revocation for PKIs and the web of trust [46] | 2020 | Generic | Bitcoin and certificate revocation Vectors | Decentralisation of revocation method for users and CAs |
| Secured surveillance storage model using blockchain [117] | 2021 | Generic | Ethereum blockchain and IPFS | Storage model with distributed capability |
| Blockchain based secure framework for IoT with user revocation [56] | 2019 | IoT | Ethereum blockchain | Blockchain-Based Secure smart contract for certificate issuance, update, revocation, and audit functions |
| Blockchain-based automated certificate revocation for 5G IoT [53] | 2020 | IoT | Elliptic Curve Qu Vanston and Hyperledger fabric blockchain | lightweight and automated certificate revocation for 5G IoT |
| Revocation mechanisms for academic certificates stored on a blockchain [122] | 2020 | Education | Blockchain, BTCerts, IPFS and SHA256 | A University diploma certificate revocation and verification |
| Application of blockchain in education: GDPR-Compliant and scalable certification and verification of academic information [34] | 2021 | Education | Private and public blockchain, merkel tree and smart contract | Academic certificate issuance and verification system |
| Privacy-preserving blockchain-based storage management [128] | 2019 | Generic | Ethereum, OCSP and CertChain | Cryptographic accumulator based distributed certificate revocation list |
| Blockchain-based certificate transparency and revocation transparency [126] | 2022 | Generic | Blockchain and X.509 PKI | Blockchain-based certificate and revocation transparency |
| An Efficient blockchain-based cross-domain authentication and secure certificate revocation scheme [49] | 2020 | Generic | Blockchain and trust certificate | Authentication and secure certificate revocation system |
| Distributed certificate revocation list [90] | 2021 | Generic | Cryptographic-accumulator, private blockchain and PKI | A Cryptographic accumulator framework based distributed certificate revocation list |
| Pseudonym certificate management scheme based on blockchain for Internet of Vehicles [20] | 2019 | IoV | Pseudonyms, blockchain, PKI | Certificate management scheme for preventing insider attacks in internet of vehicles |
| Revocable and offline-verifiable SSI [6] | 2020 | Generic | Consensus protocol, multi-signatures, bulletproofs and NIZK | Offline authentication and revocation-verification framework |
| Blockchain-based root of trust management in security credential for vehicular communications [108] | 2021 | IoV | Hyperledger Fabric, distributed PKI, root and elector certificate | Distributed PKI to protect vehicular networking privacy against an honest-but-curious authority |
| Certificate revocation management [7] | 2021 | Generic | Namecoin blockchain, bloom filter, X.509 extension certificate | Certificates revocation management and status verification system |

TABLE IV: Literature works addressing RQ3

registers which are immutable [69], which jeopardises the envisaged benefits of SSI.

*1) Revocation management:* In [7], a solution to address the scalability issue of revocation lists using blockchain registers is proposed. The revocation records that reside on CRLs are stored in a bloom filter, each filter representing a distribution with status verification. Also, the authors in [108] introduced a secured credential management system for vehicular communications that rely on blockchain for trust management, accountability, and transparency on two distinct certificates (root and elector certificates).

Additionally, the works in [47][46] proposed two different solutions for certificate revocation and management issues. In [47], the authors proposed solution to address the vulnerabilities and compromises on Registration Authorities (RA) and CAs that leads to impersonation and issuance of bogus certificates. They used a blockchain-based PKI framework which controls the activities of the RA and CA to guarantee that only trusted CAs and RAs could develop smart contracts and conduct transactions by publishing certificates in the blockchain. An RA asserts every request for a certificate and forwards it to a CA through a secure communication channel. Using a smart contract certificate transaction, an RA can then publish certificates to a blockchain network that is involved in the verification process. In [46], the authors proposed a fast blockchain-based certificate revocation for PKIs and Web of Trust (WoT). The solution leverages on blockchain technology (Bitcoin and Ethereum) to provide safe, transparent, efficient, and decentralised certificate revocation for X.509 certificates and Open PGP (Pretty Good Privacy) keys, used in the context of WoT.

Further, the authors in [117], acknowledged issues related to capturing users' data on IoT and surveillance devices. They propose a distributed and secured surveillance framework where a set of interrogation entities captures surveillance data. The framework relies on ethereum blockchain to hold the hashed value of surveillance data that are encrypted and stored on an IPFS network. With the stored data, a CA generates certificates for users, this method also addresses security challenges in IoT devices such as centralised authentication, low power device and lightweight cryptographic framework, and privacy concerns. The authors in [56] proposed a blockchain based secure framework for the IoTs with user revocation using a smart contract that supports certificate issuance, update, revocation, and audit functions that rely on an authorisation server. With the aim of addressing certificate storage over-head in IoT devices, a blockchain-based automated certificate revocation for 5G IoT is proposed in [53]. The system uses Elliptic Curve Qu Vanstone (ECQV) certificates, which are small and suitable for resource-constrained IoT devices. A smart contract using hyper-ledger fabric is used to manage certificate-related processes, such as certificate issuance and threat scoring method that automatically revokes certificates.

*2) Validation efficiency and distribution:* Traditionally, acquiring a diploma or University certificate signifies a student has completed the pre-requisite of the certificate awarded. Sometimes, students might lose access to their certificate or contact with the issuing institution, making attestation of their abilities and expertise more challenging. To address this issue, the authors in [122] extended the work in [123] to propose a framework for issuing and revoking academic certificates that are wrongly issued or misused, while storing the revocation information on a blockchain. A similar method for academic certificate verification is proposed in [34]. The solution implements public and private blockchain methods for issuing, storing, recovering, sharing, and verifying heterogeneous and unrestricted types of academic information using blockchain. Without modifying proprietary information systems, educational actors can generate and tamper-proof any type of educational data, formal or informal, and send it to a student automatically and securely whenever it is requested. This allows the students to share full or part of their certificate properties with a third party who can seamlessly check its authenticity. Issues with certificate status availability and network latency to synchronise log servers for certificates is discussed in [128]. In order to address these problems, the authors proposed a privacy-preserving blockchain-based certificate status validation mechanism (PBCert). Their approach separated revoked certificates control and storage area, where only minimal control information such as certificate hashes and related operation block height is stored on the blockchain. To preserve the privacy of the certificate owner, an external data storage is used to store detailed information about all revoked certificates, which provides an obscure response to the clients' certificate status query.

A method for certificate revocation transparency is proposed in [126] to protect and re-enforce the security of CAs. To achieve their goal, a certificate owner or web server publishes the CA-signed certificates and associated revocation status information of an SSL/TLS web server as a transaction in the global certificate blockchain. The certificate blockchain serves as an append-only public log for CAs to monitor certificate signing and revocation, and an SSL/TLS web server is all operative management over its certificates. A browser checks the certificate acquired during SSL/TLS negotiations to those in the public certificate blockchain and accepts it only if it is published and has not been revoked.

Low authentication efficiency and update gap between CRL and OCSP are common issues for cross-domain authentication using PKI infrastructure [132]. The vulnerability in these issues exposes certificates to attacks such as DDoS. In order to address the issues and increase performance, the work in [49] proposed a blockchain-based certificate for efficient cross-domain authentication and revocation using a consensus algorithm that is based on sharing the hash of random numbers. In the authors view, this method possesses the security properties of non-repudiation, anonymity, and anti-DDoS. In [90], the authors proposed a reliable revoked certificate distribution system that uses asymmetric cryptographic accumulator to condense revocation listings into a digest that can be used to validate a certificate's validity. The uses permissioned-blockchain to enhance the availability of revocation information, while lowering the danger of internal risks generated by a hacked node in the distribution system.

The enormous data generated by entities in Internet of Vehicles (IoV) when communicating between vehicle-to-vehicle, vehicle-to-infrastructure, and other networks has necessitated the need for a platform to manage data generated and transferred among the entities. Using pseudonym certificates is a common approach to solve privacy concerns of communicating entities [17]. Also, certificate management schemes are regarded as a viable approach to maintain certificate lifespan. To further preserve privacy of data shared on the devices, reduce communication overhead and simplify management of certificates, the work in [20] proposed an efficient pseudonym certificate management system in IoV. The system uses a blockchain network for reuse of pseudonym certificates, decreases and maintains the distributed CRL structure.

Identity proofing with VC requires a certificate owner to present their certificate to a verifier. For acceptance, the verifier needs up-to-date information on whether the provided credentials are still valid or have been revoked. To exchange and verify the VC, both the VC owner and the verifier needs to have an online presence. In a case where the verifier is unable to get online a verification process comes to a standstill. To address such a situation, authors in [6] extended SSI framework using zero knowledge proof to give VC owner the ability to generate a period attestation and validity status of their VC, which they can issue to a verifier for offline verification. The work uses time-based attestations of validity, which are issued by the SSI network for credentials that have not been revoked or added to a revocation list.

*3) Existing practical challenges:* In PKI, certificates are easily issued, updated, or withdrawn by the issuer without recourse to the data owner. For instance, despite GDPR's requirement that any update, withdrawal, or processing of personal data should be done with the owner's consent, a doctor's license may be revoked without his permission. Additionally, when VCs are stored on blockchain, updating personal data can be challenging as blockchain data is immutable. Moreover, since the data owner has to have prior notice of all processes on their VC and personal data, this delays service delivery and effectiveness. We further describe limitations in current works and give possible future directions.

1) Third party trust: Although some works such as in [7][46]
   [90] [122][47][53] provide a clear certificate status update method, the frameworks depends on CAs where certificate status are all downloaded to create Lightweight Revocation Status Information (LRSI) and feed filters. These systems also offer the certificate owner little to no involvement throughout the revocation process.

Although, the solution in [128] receives input from certificate owner, however the process is still executed by a CA that is fallible to SPOF. A fully sovereign identity will be defined by a system that enables certificate owners to manage every phase of the certificate's lifecycle.

2) Data breach and collusion: Distributed storage of surveillance data proposed in [117] involves collaboration between many entities. The encryption, certification and storage methods overly depends on third-parties, who can undermine the security. A failure in any of them would result in malfunction of the framework. For instance, the initial team receives data in their raw format before encryption, such entities can collude to get personal data transferred across the network. Similar threat can occur in [20], where certificate pseudonym are re-collated in a central CRL or certificate fingerprint in a blockchain [46].

3) User privacy and monitoring: In [56], the framework completely relies on an authorisation server that brokers access request, certificate status and revocation request to a blockchain register. Likewise, the solution in [34] provides a clean method for issuance and management of verifiable academic certificates on a public blockchain. However, majority of these processes do not explicitly inform data owners on the revocation procedure or notification beforehand.

### E. Addressing RQ4

For this section, the fourth search query, "delegation using blockchain," is considered. In accordance with data subjects right under GDPR guardianship and delegation, we divided retrieved literatures into the following categories: (1) delegation Management, and (2) control and transparent delegation. The delegation of control is similar to the power of attorney. It is a process of entrusting or transferring acting powers to a legal entity to act on behalf of another entity in conducting transactions. The GDPR article 80, provides for this right to data subject provided the delegatee is recognised by the law of the land. Traditionally, access delegation is achieved by authentication and authorisation processes. Authenticating a digital identity implies proving that they are indeed who they claim to be by presenting what they have (an identity document) or who they are (biometrics) [14]. While authorisation indicates a decision on what resources an authenticated digital identity is allowed to perform. Some standard protocols used in achieving these processes include; SAML [15], OAuth [103] and ABAC. A drawback of these methods is the centralisation of control to a single service, which monitors the delegation process. Therefore, if the central delegation control method fails, no permission or delegation will occur.

The ability to delegate access rights is important as it improves the quality and timeliness of service delivery. This process is mostly felt in sectors like health, banking and finance, and educational institutions. Think of an attorney, accountant, or stockbroker representing his clients in carrying out business transactions, or a physician acting a health institution. For a physician, this might mean a representation of both the healthcare institution and the patient, granting the practitioner unrestricted access to the patient's institutional and personal health record, which raises concerns on data privacy [101].

Regarding guardianship, the EU states and GDPR [96] recognise parents to individually or jointly represent their child below the age of 13. Here, parenthood has to be established typically by matching names on a certified document (national card, passport, birth certificate) of the parent to that of the child. Likewise, to prove affiliation to a health, governmental, or business institution, an entity presents an identity document issued by the institution. In **Figure** 9, the controller (parent or guardian) creates a key pair and the DID which is used to identify an under age child. This representation is well recognised and accommodated under most circumstances. As mentioned in GDPR's article 8, "a controller shall make reasonable efforts to verify in such cases that consent is given or authorised by the holder of parental responsibility for the child", taking into consideration available technology [96].

Once a child reaches maturity, they will create new key pairs and issue a request to the controller who publishes the pair. Henceforth, the child takes over control based on the new key pair and DID generated. Such examples can be used for any form of delegated representation in organisations, things, data structures and others. However the key pair in this process can be compromised. SSI addresses these using decentralised key management (DKMS) offline and social recovery methods [127]. Offline recovery is done to a data storage medium and is kept under the controller's care in a safe zone. However, risks attached to this method include: they can be misplaced, stolen, or affected by natural disasters. Social recovery uses a DKMS agent that cryptographically splits the key pairs into pieces which are then coded to be shared with trustees of the holder's choice (parent, friends, spouse e.t.c) [33]. For example, a controller can decide to share pieces of his private key with ten trustees but would only need to recover five to recover their private key. With social recovery, a holder can add or remove trustees at will. Although the DKMS addresses rotation and recovery of key pairs, it does not cover revocation of VCs. Assuming the custodian of a child gets deceased, how does a new guardian get a hold of VCs issued on behalf of the child, and possibly revoke them? We outline the proposed works of this section **Figure** 8 and **Table** VI.

*1) Delegation Management:* To prevent centralised replay attacks, cross-site request forgery, and other linked attacks, the authors in [131] proposed a distributed account management system. The system is based on a consortium of delegated blockchain architectures, ABAC and resource server among federated nodes for third-party authentication and authorisation. A proof of data ownership is proposed in [98], where personal data content on IPFS are linked anonymously with a secret code. A delegator can delegate the code with a request for the erasure of personal data on the network, which is replicated across all other nodes on the network.

The work in [70] proposes a framework to address the drawbacks of centralised access control delegation discussed in [18], where users are unable to manage authorisation policies or even keep track of access logs. The system uses a user-managed access (UMA) delegation schema with blockchain to alleviate the reliance on centralised delegation. The schema enables data owners to define access policies and maintain authorisation using smart contracts on a blockchain. The authors claimed their solution is transparent, since all entities in the network schema can interact with the smart contract directly by joining the blockchain.

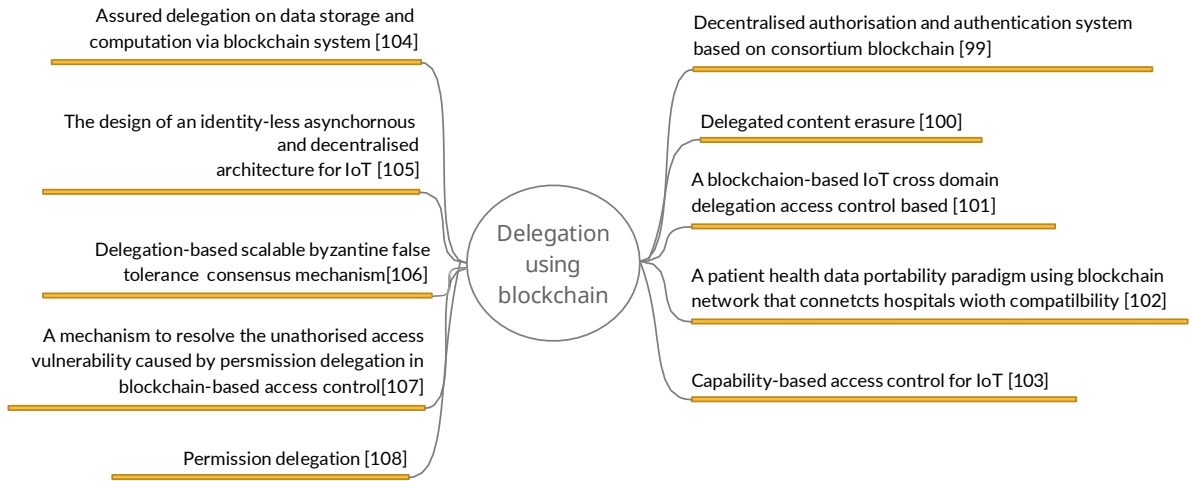

Fig. 8: Featured works in blockchain, guardianship and delegation

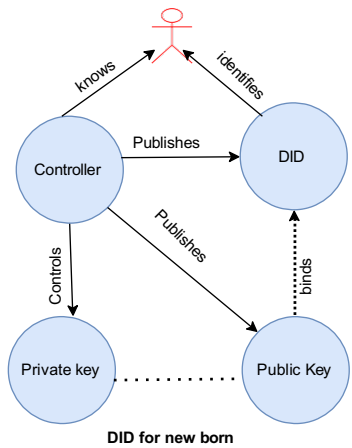

Fig. 9: Description of DID creation for data access delegation and transfer of guardianship.

Currently, patients electronic health records (EHRs) are mostly stored in siloed system/institutions holding the data. These has created inconveniences both to patients and healthcare service providers. Patients visiting multiple health centres would have to manage different accounts/health records that are fragmented and difficult to correlate when needed for treatments, follow-up and monitoring [93]. To address this issue, the authors in [106] proposed a patient-centric portable health system that is decentralised for managing patients' health data using permissioned blockchain. The system assigns sister hospitals the responsibility of managing and brokering health records owned, controlled, and managed by a health institution. It also uses cryptographic hashing techniques to anchor large amounts of data to the blockchain and store fingerprints of the data off-chain for data provenance. A capability based access control system for IoT devices is proposed in [87]. The system uses ethereum smart contracts on a blockchain to store and manage capability tokens that store and keep track of users authorised activities on specific data. In their perception, the method provides fine-grained access control and more flexible token management by defining capability tokens in units of action and using a delegation graph to preserve the token delegation relationship among the users.

*2) Control and transparent delegation:* The advancement in Internet era offers cloud computing features such as infinite outsourced storage and processing capabilities.

Despite the benefits, this setup has some problems which could affect data security and effectiveness of services such as: lack of trust, breaches in outsourced data, service reliability, and network latency [92]. Users are concerned about whether or not outsourced data can be updated or erased in remote storage as well as whether the cloud can do their intended computer tasks. To address SPOF in centralised delegation, the work in [11] proposed a trusted IoT-based framework which is decentralised and consensual using blockchain that stores all delegated IoT devices and a manager that has access to hashed values of the IoT devices in the blockchain to verify a delegations. Additionally, the authors in [95] proposed an architecture with efficient and transparent verifiable delegation for outsourced data storage and computing. The system involves data owner who outsources encrypted data to a distributed storage and publishes pointers such as abstract and tags of the data to a blockchain. Requestors will then have to scan the blockchain to locate desired data and use a policy decision point (PDP) for data integrity check. If requestors are confirmed, they contact the smart contract to apply for data utilisation. The work in [91] uses a capability-based access control method to propose an identity-less, asynchronous, and decentralised delegation model for IoT based on blockchain technology. The solution uses a private on-chain network of attribute providers, while the resources to be delegated are stored off-chain. Delegation process is brokered to a delegate on a chain of networks which is monitored by a resources manager.

The efficiency and effectiveness of blockchain services to provide secured system with high data throughput, scalability, speed of consensus confirmation and security against maliciously behaving nodes largely depend on the adopted consensus network [43]. In order to achieve a system with high scalability, the authors in [73] proposed a hybrid and delegated consensus method consisting of 5 committees for: epoch creation, confirmation, consensus, network, and campaign. The system combines Proof of Work (PoW) and Byzantine False Tolerant (BFT) to achieve a good balance of efficiency and scalability for processes within the committees. To address vulnerability issues in data management with resource controllers, authors in [115] proposed a Token-constrained Permission Delegation Algorithm (TCPDA) using blockchain and ABAC. It converts access control policy defined with ABAC into constraints for permission. This embeds the constraints in the permission token that are stored in a blockchain, and forms constraints on transferring tokens. To prevent unauthorised access or delegation, a token can only be given to digital identity that fulfils requirements of the constraints.

*3) Existing practical challenges:* The process of delegation and issuance of delegation certificates to a delegatee follow a similar pattern to issuing VC. A delegatee presents personal attributes to prepare a delegation certificate, but has little or no idea how the delegator manages the attributes. The delegator who prepares a delegation certificate knows the verifier that accepts the certificate. Therefore, even if the certificate is destroyed after usage, both the verifier and delegator can collude and perpetuate an attack by tracing delegate, tracing access pattern, profiling, or monitor earlier transactions. Other limitations found in the literature include:

1) Third party trust: The reliance on third parties, who may be complicit in the management of IoTs in blockchain and anchor trust to other health institutions, is a significant disadvantage of framework in [11][106]. Full decentralisation with no third party trust will serve better.
2) Non-flexibility and mutability: Delegated approaches proposed in [115] use access control methods on blockchain that are immutable. This results in a non-flexible approach which also contradicts GDPRs data subject's right to be forgotten as transactions can neither be changed nor completely removed from the network.
3) Single-point of failure: Delegating and outsourcing encrypted data as used in [95] could suffer from the security flaws of a centralised system. Although personal data can be protected using encryption, a malicious insider or an entity that is honest but curious with access to the ciphertext can circumvent or breach the trust and launch an attack on the data [62][78].
4) Data minimisation: Although the work in [91] proposes a fine-grained delegation within entities using capability token, the use of non-interoperable blockchain smart contracts that contain delegatee and delegator's information affects users' right to delete personal information when they desire to. Likewise, the solution does not provide a data minimisation method for precise data authorisation method.
5) Completeness and delegation management verification: Although the work in [98] assumes that a data delegation request is authenticated, integrity is checked, as well as validity of a request. There is no means to determine the completeness of the request as to whether it has been rightly executed across all nodes. Also, the deletion request is only applicable to online blocks, nodes that have acquired blocks prior to the request and never come online can continue to use such data without having to obey the deletion request.

## VI. THE GDPR PRIVACY PRINCIPLES

Though there have been no definitive criteria on how to evaluate SSI systems, several proposals have been forwarded. Authors in [58] listed some factors that affects the adoption of blockchain technology such as institutional factors, regulations and legislation, governance among others. The frameworks in [119][83] provided a broad approach that encompass data security, integrity and privacy consideration in SSI systems. However, as the SSI field continues to evolve, a holistic principle covering all SSI components is required. Faced with this, we adopted the data privacy principle outlined in GDPR article 5, 20 and others, which gave a comprehensive approach to protecting the use of personal data. According to the selected principles, we describe the required properties to maintain data owners privacy, and use these properties to evaluate earlier discussed literatures in **Table** VII, and a summary distribution in **Figure** 10.

1) **Lawfullness, fairness and transparency:** Frameworks' methods, techniques, or procedures for analysing (any process) user data must be known, lawful, open-source, and independent of any specific architecture with fairness.
2) **Purpose limitation:** Personal data must be collected for clear, explicit, and legal purposes and must not be processed in a way that is incompatible with those purposes.
3) **Data minimisation:** The disclosure of data to accomplish a task or verify claims should be as precise as possible. For example, if a person's address is to be verified, such verification should request a claimer to provide suggested address from which the service should return a boolean response (true or false) after comparison with an actual address rather than disclosing the full address of the person.
4) **Data accuracy:** Personal data must be accurate and, when required, kept up to date. Every reasonable effort must be made to ensure that personal data that is erroneous, given the purposes for which they are processed, is immediately destroyed or updated to reflect correctness.
5) **Storage limitation:** If personal information is processed purely for use in the public interest, it may be stored

| Literature description | Year | Domain | Technology | Contribution |
|---|---|---|---|---|
| Decentralised authorisation and authentication system based on consortium blockchain [131] | 2021 | Generic | Byzantine false tolerance, ABAC and Ethereum | A consortium blockchain architecture and designs protocols for account management and distributed consensus |
| Delegated content erasure [98] | 2020 | Generic | IPFS | Anonymous protocol for delegated content erasure requests in the IPFS |
| A blockchain-based IoT cross-domain delegation access control method [70] | 2021 | IoT | Smart contracts and UMA protocol | Multi domain delegation trajectory aggregation that supports forensic analysis of intra/cross-domain delegation |
| A patient health data portability paradigm using blockchain network that connects hospitals with compatibility [106] | 2020 | Healthcare | Hyperledger fabric and decentralised model | Patient Data Portability on the Cloud Utilising Delegated Identity Management |
| Capability-based access control for IoT [87] | 2019 | IoT | Ethereum smart contract and Capability-based access control | Distributed access control, delegation and verification method for the IoT |
| Assured delegation on data storage and computation via blockchain System [95] | 2019 | Generic | Verifiable computation, proxy re-encryption and smart contract | An efficient and transparent public verifiable delegation for both storage and computing |
| The design of an identity-less, asynchronous, and decentralised delegation architecture for IoT [91] | 2019 | IoT | Capability based access control and ethereum | Delegation Model for the IoT |
| Delegation based scalable byzantine false tolerance consensus mechanism [73] | 2020 | Generic | Byzantine false tolerant and proof of work | Delegation based fault tolerance mechanism |
| A Mechanism to resolve the unauthorized access vulnerability caused by permission delegation in blockchain-based access control [115] | 2021 | Generic | Permission delegation algorithm, RBAC and ethereum blockchain | Token-constrained permission delegation mechanism |
| Permission delegation [11] | 2019 | IoT | Blockchain and delegated permission agent | Access control with permission delegation in IoT |

TABLE V: Literature works addressing RQ4

longer or in a format that prevents the owner from being identified.

6) **Integrity and confidentiality:** Personal data must be treated in a way that ensures proper security, including safeguards against unauthorized use. In situations where data usage in service conflicts with rights of data subject, personal rights should prevail and personal data should be protected.
7) **Accountability:** Data controllers and organisations handling digital data are to assume accountability for adhering to GDPR principles of processing personal data and provide evidence that they do so.

VII. CRITICAL EVALUATION AND DISCUSSION

Although the first blockchain implementation is used for financial transactions in bitcoin [85], its wide adoption by SSI has gone far beyond that. Many advantages provided by blockchain and VC have created an ecosystem that is suitable in a wide variety of industry services including government, banking and finance, education, and healthcare. Although, most SSI solutions have been tied to using blockchain, SSI is not blockchain-dependent [48]. The authors in [28] noted some user-centric SSI concepts that do not require blockchain. Likewise, several literature have achieved SSI goal of decentralisation and some properties of the GDPR principles without the use of blockchain.

We evaluated some selected literatures in **Table** VII, using the adopted GDPR data privacy principles. These include both blockchain and non-blockchain solutions. Substantially, some literature fulfils basic properties of the evaluation principle such as data integrity and transparency. However, the design concepts of some literatures provided little or no methods to support data minimisation and purpose limitation of personal data. Additionally, despite the success attributed to blockchain adoption in some literature, there exists a number of limitations and problems that needs to be considered [72][2].

1) **Technical issues**: In SSI, service and users portability and interoperability are crucial to their sovereignty, yet most literature and technologies only support the portability of users within their network or similar networks of architectural design. Because blockchain network is a new technology that is growing, they have continued to expand. Whether these technologies will survive the test of time requires a lot of attention from governments, researchers, and industry organisations. As of the time of writing this document, there are about a hundred registered siloed, independent blockchain networks with W3C that are unconnected[32]. Each blockchain comprises varying features, protocols and standards causing fragmentation in adoption [112], hence interoperation between blockchains only occurs among technologies with same open source implementation or using a trusted party [22]. Although GDPR's article 20, provides data subject's right to collecting, sharing, limiting access, and interoperating personal data seamlessly as they wish with no hindrance, current blockchain solutions do not provide direct interoperability without a trusted third party [67]. These jeopardises the foreseen data privacy principles of GDPR and SSI. Also, the complexity in underlying blockchain network protocols results in strenuous user interfaces[55].

| Literature description | Year | Domain | Technology | Contribution |
|---|---|---|---|---|
| Decentralised authorisation and authentication system based on consortium blockchain [131] | 2021 | Generic | Byzantine false tolerance, ABAC and Ethereum | A consortium blockchain architecture and designs protocols for account management and distributed consensus |
| Delegated content erasure [98] | 2020 | Generic | IPFS | Anonymous protocol for delegated content erasure requests in the IPFS |
| A blockchain-based IoT cross-domain delegation access control method [70] | 2021 | IoT | Smart contracts and UMA protocol | Multi domain delegation trajectory aggregation that supports forensic analysis of intra/cross-domain delegation |
| A patient health data portability paradigm using blockchain network that connects hospitals with compatibility [106] | 2020 | Healthcare | Hyperledger fabric and decentralised model | Patient Data Portability on the Cloud Utilising Delegated Identity Management |
| Capability-based access control for IoT [87] | 2019 | IoT | Ethereum smart contract and Capability-based access control | Distributed access control, delegation and verification method for the IoT |
| Assured delegation on data storage and computation via blockchain System [95] | 2019 | Generic | Verifiable computation, proxy re-encryption and smart contract | An efficient and transparent public verifiable delegation for both storage and computing |
| The design of an identity-less, asynchronous, and decentralised delegation architecture for IoT [91] | 2019 | IoT | Capability based access control and ethereum | Delegation Model for the IoT |
| Delegation based scalable byzantine false tolerance consensus mechanism [73] | 2020 | Generic | Byzantine false tolerant and proof of work | Delegation based fault tolerance mechanism |
| A Mechanism to resolve the unauthorized access vulnerability caused by permission delegation in blockchain-based access control [115] | 2021 | Generic | Permission delegation algorithm, RBAC and ethereum blockchain | Token-constrained permission delegation mechanism |
| Permission delegation [11] | 2019 | IoT | Blockchain and delegated permission agent | Access control with permission delegation in IoT |

TABLE VI: Literature works addressing RQ4

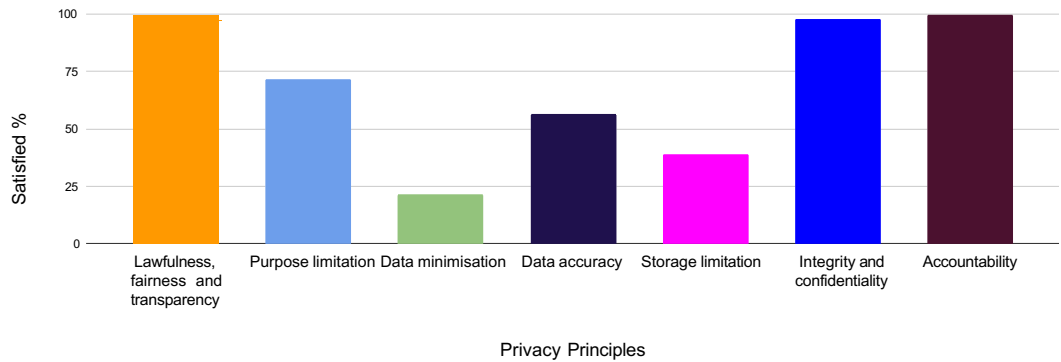

Fig. 10: A summary distribution of GDPR privacy principles

2) **Non-flexible digital wallets**: So far, mobile wallets (which contain bank cards, travel papers, and so on) and cryptocurrency wallets (which are used for cryptocurrency commerce in bitcoin, and so on) have been major developers of digital wallets. Because the wallets may store a digital certificate and release their attributes for authentication by the holder as needed, either of them should function perfectly with the required structure of SSI. However, their implementation threatens the sovereignty of VC holders who will have to depend on third-party applications that resides at different locations or even requires them to have some prior technical know-how before usage. Also, a good number of the networks do not support cross-chain interoperability transactions. A self-sovereign digital wallet should be built on open standards, with the ability to accept any kind of standardised VC. It should also be open to devising portability and be scalable enough to accommodate as many VCs as desired by the holder.

| References | Privacy Principles | | | | | | |
|---|---|---|---|---|---|---|---|
| | Lawfulness, fairness and transparency | Purpose limitation | Data minimisation | Data accuracy | Storage limitation | Integrity and confidentiality | Accountability |
| Awuson et al [19] | ✓ | ✓ | ✗ | ✗ | ✗ | ✓ | ✓ |
| Souza et al [118] | ✓ | ✗ | ✓ | – | ✗ | ✓ | ✓ |
| Sharma et al [113] | ✓ | ✗ | ✓ | ✓ | ✗ | ✓ | ✓ |
| Barati et al [21] | ✓ | ✓ | ✓ | ✗ | ✗ | ✓ | ✓ |
| Nguyen et al [120] | ✓ | ✗ | ✗ | ✓ | ✗ | ✓ | ✓ |
| Imran et al [76] | ✓ | ✗ | ✗ | ✓ | ✓ | ✓ | ✓ |
| Rantos et al [102] | ✓ | ✓ | ✗ | ✗ | ✗ | ✓ | ✓ |
| David et al [51] | ✓ | ✓ | ✗ | – | ✗ | ✓ | ✓ |
| Maryam et al [31] | ✓ | ✓ | ✓ | ✗ | ✓ | ✓ | ✓ |
| Mirko et al [66] | ✓ | ✓ | ✗ | ✗ | ✗ | ✓ | ✓ |
| Edman et al [109] | ✓ | ✗ | ✗ | ✗ | ✗ | ✓ | ✓ |
| Esmel et al [30] | ✓ | ✓ | ✗ | ✓ | ✓ | ✓ | ✓ |
| Bayat et al [99] | ✓ | ✗ | ✗ | ✗ | ✓ | ✓ | ✓ |
| Jamila et al [12] | ✓ | ✗ | ✗ | ✗ | ✓ | ✓ | ✓ |
| Nudge et al [125] | ✓ | ✓ | ✗ | ✓ | ✗ | ✓ | ✓ |
| Pedrosa et al [94] | ✓ | ✓ | ✓ | – | ✗ | ✓ | ✓ |
| Dominik et al [111] | ✓ | ✓ | ✗ | ✗ | ✗ | ✓ | ✓ |
| Davidchadwick et al [27] | ✓ | ✓ | ✗ | – | ✗ | ✓ | ✓ |
| Amal et al [5] | ✓ | – | ✗ | ✓ | ✓ | ✓ | ✓ |
| Deniz et al [107] | ✓ | ✓ | ✗ | – | ✓ | ✓ | ✓ |
| Fu et al [45] | ✓ | ✓ | ✗ | ✓ | – | ✓ | ✓ |
| Peter et al [54] | ✓ | ✗ | ✗ | ✓ | ✓ | ✓ | ✓ |
| Adja et al [7] | ✓ | ✓ | ✗ | ✓ | ✗ | ✓ | ✓ |
| Chang et al [108] | ✓ | ✓ | ✗ | – | – | ✓ | ✓ |
| Ramana et al [117] | ✓ | ✗ | ✗ | ✓ | ✗ | ✓ | ✓ |
| Hu et al [56] | ✓ | ✓ | ✗ | ✓ | ✗ | ✓ | ✓ |
| Gu et al [49] | ✓ | ✓ | ✓ | ✓ | – | ✓ | ✓ |
| Abba et al [47] | ✓ | ✗ | ✗ | ✓ | ✗ | ✓ | ✓ |
| Vidal et al [122] | ✓ | ✗ | ✗ | ✓ | ✓ | ✓ | ✓ |
| Garba et al [46] | ✓ | ✓ | ✗ | – | ✓ | ✓ | ✓ |
| Ozcelik et al [90] | ✓ | ✓ | ✗ | ✓ | – | ✓ | ✓ |
| Hewa et al [53] | ✓ | ✓ | ✗ | ✓ | ✓ | ✓ | ✓ |
| Yao et al [128] | ✓ | ✓ | ✗ | ✓ | ✓ | ✓ | ✓ |
| Ze et al [126] | ✓ | ✓ | – | – | ✗ | ✓ | ✓ |
| Bao et al [20] | ✓ | ✓ | ✗ | ✓ | – | ✓ | ✓ |
| Abraham et al [6] | ✓ | – | ✓ | – | ✗ | ✓ | ✓ |
| Politou et al [98] | ✓ | ✓ | ✓ | – | ✗ | – | ✓ |
| Liao et al [70] | ✓ | ✗ | ✗ | ✓ | ✗ | ✓ | ✓ |
| Pal et al [91] | ✓ | ✓ | ✗ | ✓ | ✗ | ✓ | ✓ |
| Sabir et al [106] | ✓ | ✓ | ✗ | ✗ | ✓ | ✓ | ✓ |
| Yuan et al [73] | ✓ | ✓ | ✗ | – | – | ✓ | ✓ |
| Jinsha et al [115] | ✓ | ✓ | ✗ | ✓ | ✓ | ✓ | ✓ |
| Asif et al [11] | ✓ | ✓ | ✗ | ✓ | – | ✓ | ✓ |
| Pei et al [95] | ✓ | ✓ | ✓ | – | ✓ | ✓ | ✓ |
| Sasabe et al [87] | ✓ | ✓ | ✗ | ✓ | ✓ | ✓ | ✓ |
| Lius et al [34] | ✓ | ✓ | ✓ | ✓ | – | ✓ | ✓ |

TABLE VII: Summary of selected literature and their compliance with GDPR privacy principles. Works that satisfy each principle is annotated with ✓, while ✗, represent unsatisfied property. Undetermined crititerias are denoted with –.

3) **Regulation and legislation**: Data transparency, immutability and smart contracts are among the many benefits of blockchain. Despite the adoption in many areas, to date, there is no acceptable global standard for blockchain adoption. Different proprietors design and build systems to address specific scenarios. For example, the registry of transactions in bitcoin differs from that of ethereum and many more. Likewise identity creation, certificate generation, utilisation, and revocation. Without a common and globally acceptable standardisation of blockchain, global acceptance and usage is still a mirage that would create more complexity rather than simplicity [58].

4) **Credential and key management**: Typically, digital wallets act as the sole container that empowers users to store their VC. Some solutions such as the use of biometrics with fingerprints for strong authentication, are used to protect it. Also, in the event of loss or compromised keys, DKMS is used for key recoveries. However, these recoveries do not cover compromised VCs or thwart any damage that might arise from such a situation when noticed. Failure to detect a revoked or compromised certificate could result in a system compromise that could be catastrophic. The work in [115] used delegated permission token for certificate management, but once a digital wallet or mobile device is compromised, such token or VC can be stolen and used to mimic a legitimate data subject's proof of identity with no method to detect the compromise.

## VIII. CONCLUSION

This study aims to identify the compliance of literature works to GDPR privacy principles through a systematic review of the most recent literature in SSI and blockchain solutions. We outlined 4 research questions and conducted an in-depth search in 3 English language peer-reviewed databases: Web of Science, Scopus, and IEEE explore. We categorised retrieved documents according to their literature approach and link them with rights of data subjects under GDPR. From the literature, we noticed blockchain is widely adopted for providing digital and online identity services, followed by IoT and smart cities, then health care, educational and financial domains. Blockchain methods for erasing certificates and personal data, and certificate revocation are the most discussed among articles retrieved. We evaluated the proposals using GDPR privacy principles. In general the literature works analysed satisfied 69.5% of all GDPR privacy principles. All the literature met by 100% principles of lawfulness, fairness and transparency, and accountability. Also, these works satisfies the principles of integrity and data confidentiality, purpose limitation, data accuracy, storage limitation and data minimisation by 97.8%, 71.7%, 56.5%, 39.1% and 21.7% respectively. Limitations of this work includes the inclusion of only peer-reviewed articles, exclusion of industry reports and thesis works. Also, our review mostly covered SSI systems that use blockchain technology with a few non-blockchain. Furthermore, the scope of this work covers a limited area of GDPR and SSI principles, and our search only included articles within two years (2019-June 2021).

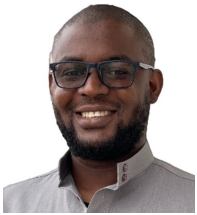

**Abubakar-Sadiq Shehu** holds a Ph.D. in Computer Science from the University of Porto, Portugal, and a Master of Science in Business Information Systems from the University of East London. He previously earned a Bachelor of Science in Computer Science (Digital System Security) from the University of Wollongong, Australia. Currently, Abubakar-Sadiq is a cybersecurity research fellow at the University of Surrey, United Kingdom. His research interests include data privacy and security, secure access and data delegation.

JOURNAL OF LATEX CLASS FILES, VOL. 14, NO. 8, AUGUST 2015 25